\documentclass[structabstract]{aa}  

\usepackage{txfonts}
\usepackage{graphicx}
\usepackage{cancel}
\usepackage{multirow,bigdelim}
\usepackage{supertabular}
\usepackage{longtable}
\usepackage{url}
\usepackage{rotating}
\usepackage{pdflscape}

\newcommand{\msun}{\ensuremath{\mathit{M}_{\odot}}}

\newcommand{\vrot}{\ensuremath{\varv_{\rm e}\sin i}}
\newcommand{\vsini}{\ensuremath{\varv_{\rm e}\sin i}}

\def\kms{\mbox{${\rm km}\:{\rm s}^{-1}$}}

\def\lesssim{\mathrel{\hbox{\rlap{\hbox{\lower4pt\hbox{$\sim$}}}\hbox{$<$}}}}

\def\gtrsim{\mathrel{\hbox{\rlap{\hbox{\lower4pt\hbox{$\sim$}}}\hbox{$>$}}}}

\defcitealias{evans2011}{Paper~I}
\defcitealias{dufton}{Paper~X}
\defcitealias{sana}{Paper~VIII}
\defcitealias{ramirezagudelo}{Paper~XII}
\defcitealias{doran}{Paper~XI}

\newcommand{\mrm}[1]{\ensuremath{\mathrm{#1}}}

\newcommand{\Teff}{\ensuremath{T_\mrm{eff}}}
\newcommand{\logL}{\ensuremath{\log\,L}}
\newcommand{\logg}{\ensuremath{\log\,\mrm{g}}}

\newcommand{\mdot}{\ensuremath{\dot{M}}}
\newcommand{\rstar}{\ensuremath{R}}
\newcommand{\vinf}{\ensuremath{\varv_{\infty}}}

\newcommand{\vmicro}{\ensuremath{\xi_\mrm{m}}}

\newcommand{\Weq}{\ensuremath{W_\mathrm{eq}}}
\newcommand{\nabun}{\ensuremath{\varepsilon_\mathrm{N}}}

\def\SNR{\ensuremath{\mathrm{S/N}}}

\def\figpath{./}
\def\hunterplotpath{./}

\titlerunning{Surface nitrogen abundances of presumably single O-type giants and supergiants}
\begin{document}
  \title{The VLT-FLAMES Tarantula Survey\thanks{Based on observations collected at the European Organisation for Astronomical Research in the Southern Hemisphere under ESO programme 182.D-0222.} 
}

   \subtitle{XXV. Surface nitrogen abundances of O-type giants and supergiants}
   \author{ 
     N.J. Grin   \inst{1,2}
     \and
     O.H. Ram\'{i}rez-Agudelo   \inst{1,2}
     \and
     A. de Koter  \inst{1,3}
     \and
     H. Sana      \inst{3}
     \and
     J. Puls \inst{4}
     \and
     I. Brott \inst{16}
     \and
     P.A. Crowther \inst{5}
     \and
     P.L. Dufton \inst{17}
     \and
     C.J. Evans \inst{6}
     \and
     G. Gr\"afener \inst{2}
     \and
     A. Herrero \inst{13,14}
     \and
     N. Langer \inst{2}
     \and
     D.J. Lennon \inst{7}
     \and
     J.Th. van Loon \inst{8}
     \and
     N. Markova \inst{15}
     \and
     S.E. de Mink \inst{1}
     \and
     F. Najarro \inst{12}
     \and
     F.R.N. Schneider \inst{9}
     \and
     W.D. Taylor \inst{6}
     \and
     F. Tramper \inst{7}
     \and
     J.S. Vink \inst{10}
     \and
     N.R. Walborn \inst{11}
}
\institute{ 
          Astronomical Institute Anton Pannekoek,
          Amsterdam University, 
          Science Park 904, 1098~XH,
          Amsterdam, The Netherlands\newline
          \email{nathgrin@gmail.com}
\and 
           Argelander-Institut f\"ur Astronomie,
           Universit\"at Bonn,
           Auf dem H\"ugel 71,
           53121 Bonn, Germany
\and 
           Institute of Astrophysics,
           KU Leuven,
           Celestijnenlaan 200D,
           3001, Leuven, Belgium
\and 
        LMU Munich, Universit\"atssternwarte,
        Scheinerstrasse 1,
        81679 M\"unchen,
        Germany          
\and 
           Departament of Physic and Astronomy
           University of Sheffield,
           Sheffield,S3 7RH,
           United Kingdom
\and 
           UK Astronomy Technology Centre,
           Royal Observatory Edinburgh,
           Blackford Hill, Edinburgh, EH9 3HJ, United Kingdom
\and 
           European Space Astronomy Centre (ESAC),
           Camino bajo del Castillo, s/n
           Urbanizacion Villafranca del Castillo,
           Villanueva de la Ca\~nada,
           E-28\,692 Madrid, Spain 
\and   
            Lennard-Jones Laboratories,
            Keele University,
            Staffordshire,
            ST5 5BG, United Kingdom   
\and 
           Department of Physics,
           University of Oxford,
           Keble Road,
           Oxford OX1 3RH,
           United Kingdom
\and 
           Armagh Observatory,
           College Hill,
           Armagh, BT61 9DG,
           Northern Ireland,
           United Kingdom 
\and  
            Space Telescope Science Institute,
            3700 San Martin Drive,
            Baltimore, MD 21218, USA 
\and  
            Centro de Astrobiolog\'{i}a (CSIC-INTA),
            Ctra. de Torrej\'on a Ajalvir km-4,
            E-28850 Torrej\'on de Ardoz,
            Madrid, Spain
\and 
           Departamento de Astrof\'{i}sica,
           Universidad de La Laguna,
           Avda. Astrof\'{i}sico Francisco S\'{a}nchez s/n,
           E-38071 La Laguna, Tenerife, Spain
\and 
           Instituto de Astrof\'{i}sica de Canarias,
           C/ V\'{i}a L\'{a}ctea s/n, E-38200 La Laguna, Tenerife,
           Spain
\and  
            Institute of Astronomy with NAO,
            Bulgarian Academy of Sciences,
            PO Box 136, 4700 Smoljan, Bulgaria
\and 
            University Vienna, 
            Department of Astrophysics, 
            T\"urkenschanzstr. 17, 1180 Vienna, Austria
\and 
                Astrophysics Research Centre,
                School of Mathematics and Physics,
                Queen's University of Belfast,
                Belfast BT7 1NN,
                United Kingdom
}

   \date{Received ....}

 
 \abstract
   {Theoretically, rotation-induced chemical mixing in massive stars has far reaching 
    evolutionary consequences, affecting the sequence of morphological phases, lifetimes, 
    nucleosynthesis, and supernova characteristics.
    }
   {Using a sample of 72 presumably single O-type giants to supergiants observed in the context of the VLT-FLAMES
    Tarantula Survey (VFTS), we aim to investigate rotational mixing in evolved core-hydrogen burning stars
    initially more massive than 15\,\msun\ by analysing their surface nitrogen
    abundances.}
   {Using stellar and wind properties derived in a previous VFTS study we computed
    synthetic spectra for a set of up to 21 N\,{\sc ii-v} lines in the optical spectral range, using the non-LTE atmosphere code {\sc fastwind}.
    We constrained the nitrogen abundance by fitting the equivalent widths of relatively strong
    lines that are sensitive to changes in the abundance of this element. Given the quality of the data, we constrained the nitrogen abundance in 38 cases; for 34 stars only upper limits could be derived, which includes almost all stars rotating at $\vrot\ >200\,\kms$. }
   {We analysed the nitrogen abundance as a function of projected rotation rate \vrot\ and confronted it with predictions  of rotational mixing. We found a group of N-enhanced
    slowly-spinning stars that is not in accordance with predictions of rotational mixing
    in single stars. Among O-type stars with (rotation-corrected) gravities less than $\log\,g_c = 3.75$ this group
    constitutes 30$-$40 percent of the population.
    We found a correlation between nitrogen and helium abundance which is consistent with expectations, suggesting that, whatever the mechanism that brings N to the surface, it displays CNO-processed material. 
    For the rapidly-spinning O-type stars we can only provide upper limits on the nitrogen abundance, which are not in violation with theoretical expectations. Hence, the data cannot be used
    to test the physics of rotation induced mixing in the regime of high spin rates.}
  {While the surface abundances of 60-70 percent of presumed single O-type giants to supergiants behave in conformity with expectations, at least 30-40 percent of our sample can not
   be understood in the current framework of rotational mixing for single stars. Even though we have excluded
   stars showing radial velocity variations, of our sample may have remained contaminated
   by post-interaction binary products. Hence,
   it is plausible that effects of binary interaction need
   to be considered to understand their surface properties. Alternatively, or in conjunction, 
   the effects of magnetic fields or alternative mass-loss recipes may need to be invoked.}
  
   \keywords{
   			stars: early-type --
             stars: abundances, rotation -- 
             Magellanic Clouds --
             Galaxies: star clusters: individual: 30 Doradus --
			line: profiles             \vspace{-3mm}
               }

   \maketitle

%

\section{Introduction}\label{sec:intro}




Despite the importance of massive stars for Galactic and extragalactic astrophysics, many of the 
physical processes that control the evolution of these objects are still not well understood
\citep[see e.g.,][]{langer2012}. 
%
As one of the key agents of massive star evolution, the effects of rotation are manifold.
The internal structure of spinning stars becomes latitude dependent \citep{zeipel1924}
and centrifugal forces resulting from rotation may lead to deviation 
from a spherical shape \citep[e.g.,][]{collins1963,townsend2004}. The von Zeipel effect
results in such stars showing relatively hot and bright polar regions and relatively cool and dim equatorial zones -- effects that are actually observed \citep{domicianodesouza2003,domicianodesouza2005}. 
Spinning stars have longer main-sequence lifetimes \citep{brott2011a,ekstrom2012,kohler2015} and may follow
different paths in the Hertzsprung-Russell diagram (HRD), as the centrifugal force 
reduces the effective gravity
and because rotation may also impact mass-loss and angular momentum loss at the surface
\citep[see][for an extensive discussion]{maeder2009,langer2012}. 

Rotation induced instabilities trigger
internal mixing, transporting material from deep layers to the surface. In extreme cases, this
mixing may be so efficient that the stars remain chemically homogeneous throughout their lives and -- because
of this -- avoid envelope expansion \citep{maeder1987}. In special cases, this type of evolution has 
been proposed to lead to long-duration gamma ray bursts 
\citep[e.g.,][]{yoon2005,woosley2006}. Chemical homogeneity is further a pivotal ingredient for certain formation channels of close-binary black holes, that may over time merge and emit a gravitational wave signal \citep{demink2009,mandel2016,marchant2016}.

Helium may, in principle, serve as a tracer of the efficiency of rotationally induced
mixing processes. However, as it is produced on the nuclear timescale, the anticipated surface enrichment of He 
is relatively minor. In this paper we focus on the surface abundance of nitrogen, which is a much more sensitive agent.
In the CN (and CNO) cycle, carbon (and oxygen) are converted into nitrogen. Processed material that can escape from the core before a chemical gradient is 
established at the core boundary, can be mixed into the envelope \citep{maedermeynet1997}. This material 
will, over time, reach the stellar surface. The faster the star is spinning, the more quickly the 
material will surface and a greater surface N abundance will be reached. 

\citet{hunter2008} searched for a correlation between surface nitrogen abundance and projected
spin velocity of a sample of evolved main-sequence early B-type stars, observed in the context of
the VLT-FLAMES Survey of Massive Stars \citep{evans2006}. A subset of their sample displayed such a
correlation and was used by \citet{brott2011b} to calibrate the efficiency of rotational mixing at the metallicity of the Large Magellanic Cloud (LMC).
However, two groups of stars did not concord with the expectations of rotational mixing. We discuss this further in Sect.~\ref{sec:otherNstudies}, comparing their findings for B stars with ours for O stars. 

Here, we determine the nitrogen abundance of LMC O-type giants and supergiants that have been 
observed in the context of the VLT-FLAMES Tarantula Survey \citep[VFTS;][]{evans2011}.
The main questions we want to address are: 
what is the behaviour of these O-type stars in terms
of N-abundance as a function of projected spin velocity? 
Does our sample also contain
groups of stars that are not in agreement with the predictions of rotational mixing in single
stars? If so, can these be identified as the counterparts of the peculiar groups identified
among the B dwarfs by \citet{hunter2008}?

The paper is organised as follows. In Sect.~\ref{sec:sample} we briefly introduce
the sample of O giants and supergiants. The method of nitrogen abundance analysis is 
explained in Sect.~\ref{sec:method}. The results are presented in 
Sect.~\ref{sec:results}, compared to population synthesis predictions in 
Sect.~\ref{sec:popsyn}, and discussed in Sect.~\ref{sec:discussion}. Finally, we summarise 
our findings in Sect.~\ref{sec:conclusions}.

\section{Sample and data}
\label{sec:sample}

The VFTS project and the data have been described in \citet{evans2011}. 
In short, it is a multi-epoch study of about 800 O- and B stars in the 30~Doradus region of the LMC.
The data analysed in this paper were obtained with the Medusa-Giraffe mode of the Fibre Large Array
Multi-Element Spectrograph instrument \citep[FLAMES;][]{pasquini2002}, mounted on the Very Large 
Telescope (VLT) at Cerro Paranal, Chile. Spectral coverage provided by the adopted settings is 
$\lambda\lambda\, 3960$--$5070\,\AA$ and $\lambda\lambda\, 6442$--$6817\,\AA$, at spectral resolving powers 
$R$\,$\sim$\,8000 and $R$\,$\sim$\,16000, respectively; 
full details are given by \citet{evans2011}.

Multi-epoch observations allowed for a selection of spectroscopic binaries on the basis of their 
radial velocity (RV) variation \citep{sana2013}. We consider that stars are spectroscopically single 
if their peak-to-peak RV variation is either statistically insignificant or significant but 
smaller than $20\,\kms$. About half of the stars in the latter category are however genuine 
spectroscopic binaries (Almeida et al. in prep; see also Sect.~\ref{sec:limitations}). Spectral classification of the O-type content
has been presented by  \citet{walborn2014}. Here we analyse the nitrogen content of the presumably single O-type stars classified by \citet{walborn2014} as giants, bright giants and supergiants. An overview of the distribution of our sample stars among luminosity classes {\sc III} to {\sc I} is given in Table~\ref{tab:sampleparams}.
These luminosity classes could be assigned in 72 cases. For 31 stars no luminosity class identifier could be given. We provide estimates for their nitrogen abundance in Appendix~\ref{app:datatablenoLC}. 

\section{Method}
\label{sec:method}

\subsection{Atmospheric parameters}
\label{sec:params}
\begin{figure}
\centering
\includegraphics[width=\linewidth]{\figpath 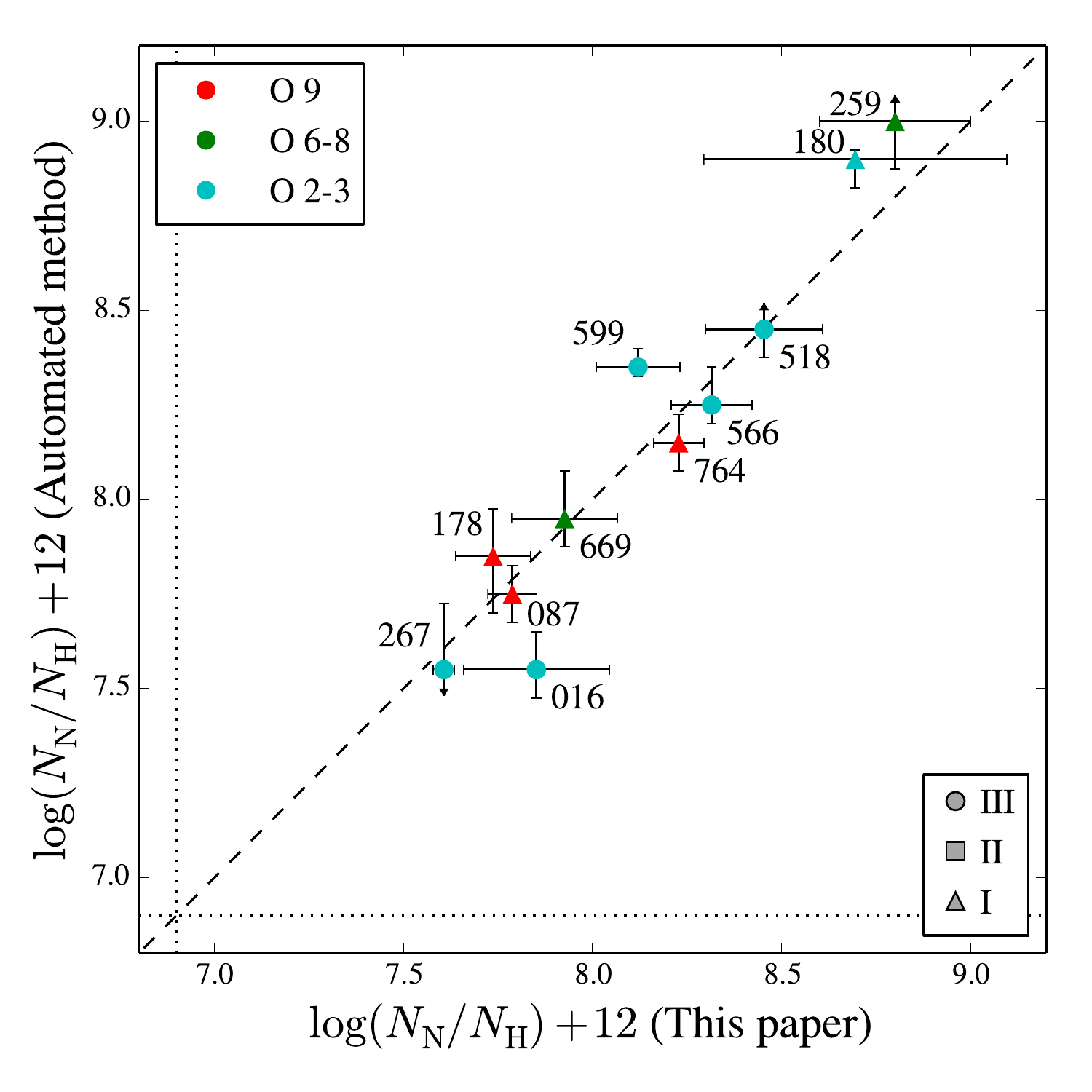}
\caption{Comparison between nitrogen abundances derived with the automated fitting method (vertical axis) 
and the method described in Sect.~\ref{sec:method} (horizontal axis). 
Dotted lines indicate the nitrogen baseline abundance of LMC stars. The dashed line shows
the 1:1 relation. The observed stars are marked with their VFTS ID. Except for two stars 
the two methods agree within 1$\sigma$. 
Note that \vmicro\ is allowed to vary in the automated method, while \vmicro\ is fixed at $10\,\kms$ for the measurements in this paper (see Sect.~\ref{sec:limitations}). }
\label{fig:GAplot}
\end{figure}

Atmospheric properties of the sample discussed here have been determined by Ram\'{i}rez-Agudelo et al. (subm.), using an automated fitting method \citep{mokiem2005,tramper2011,tramper2014}. In short, the method combines an analysis of normalised spectra using the non-LTE stellar atmosphere model {\sc fastwind}\footnote{In this work, version 10.1/7.2 of {\sc fastwind} is used.} \citep{puls2005}  with the genetic fitting algorithm {\sc pikaia} \citep{charbonneau1995}. {\sc fastwind} accounts for a trans-sonic, stationary, radial stellar outflow. 
Six stellar properties have been determined: effective temperature \Teff, surface gravity $g$, (unclumped) mass-loss rate \mdot, helium content $N_{\rm He}$, 
a depth-independent micro-turbulent velocity \vmicro, and projected equatorial rotational velocity \vrot.
The luminosity $L$ and radius \rstar\ of the star were constrained using the self-consistently computed $K_\mathrm{S}$ bolometric corrections and the Stefan-Boltzmann law.
The stellar wind was assumed to accelerate following a $\beta$-type velocity law, which is prescribed
by the flow acceleration parameter $\beta$ and the terminal flow velocity $\vinf$. Appropriate
values for $\beta$ are adopted from hydrodynamical simulations \citep{muijres2012}; values
for $\vinf$ follow from a scaling relation with the surface escape velocity \citep{kudritzki2000,leitherer1992}.

For relatively hot stars, an accurate estimate of the effective temperature from H and He lines only is severely compromised when the 
He\,{\sc i} lines disappear from the spectrum; for relatively cool stars when He\,{\sc ii} fades away. 
Because of this, Ram\'{i}rez-Agudelo et al. (subm.) have added nitrogen to their 
set of hydrogen and helium diagnostic lines for a total of 11 stars (see Fig.~\ref{fig:GAplot}).
The inclusion of nitrogen not only enables better constraints on \Teff, based on the corresponding ionisation equilibrium, but also allows to derive nitrogen abundances in parallel.
As will be discussed in Sect.~\ref{sec:abundance}, we adopt an alternative, faster method to determine the nitrogen fraction of all stars in our sample.
In Fig.~\ref{fig:GAplot}, we compare the nitrogen abundance determined by the automated method with our own measurements for the 11 sources mentioned.
Save for two sources that deviate by two sigma, both methods agree within one standard deviation. This level of compatibility is in agreement with statistical fluctuations, that are expected for independent measurements in a sample of 11 sources.   
For the sake of consistency, we adopt the abundances as measured by the method presented in Sect.~\ref{sec:abundance} throughout this work.

\subsection{Nitrogen diagnostic lines}
\label{sec:N_spectroscopy}

The physics of nitrogen line-formation and its implementation into {\sc fastwind} have been extensively described and tested by \citet{riverogonzalez2011,riverogonzalez2012b,riverogonzalez2012a}. Here we discuss the diagnostic lines used in the work at hand.

\begin{figure}
\centering
\includegraphics[scale=0.385]{\figpath 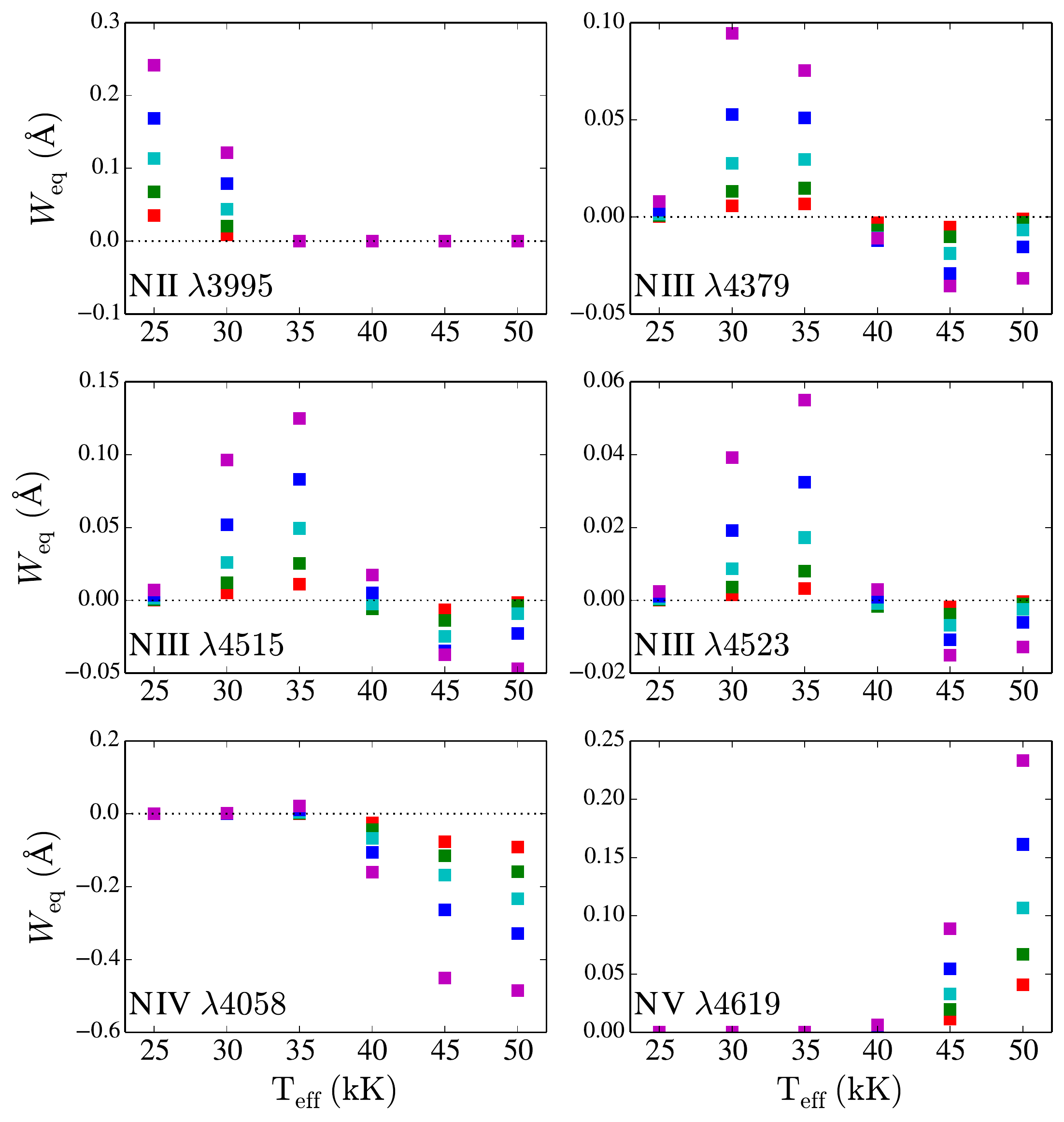}
\caption{Equivalent width as a function of effective temperature and nitrogen abundance 
for several diagnostic lines. Colours, following the rainbow from red to purple, indicate nitrogen abundances \nabun\ 
ranging from 6.93 to 8.53 in steps of 0.4 dex. Negative values of \Weq\ represent emission. For all models: $\logg=3.9\,(\mathrm{cgs})$, 
$\log \dot{M} = -6\,(\msun/\mathrm{yr})$ and $\vmicro=7\,\kms$.}
\label{fig:TMNgrid}
\end{figure}

In the optical spectral range available to us (covering $\lambda\lambda\, 3960$$-$$5070\,\AA$ and $\lambda\lambda\, 6442$$-$$6817\,\AA$) several tens of nitrogen lines can be identified. Not all of these lines are suited for spectral analysis. To select the lines that are most suitable for our purposes we applied the following selection criteria:
 {\em (i)} the lines should not be blended; {\em (ii)} the lines should be strong enough to be detectable given the typical signal-to-noise of our spectra and the range of projected spin velocities covered by our stars; {\em (iii)}  the line strengths should be as sensitive as possible to changes in abundance. 
To assess the latter we computed a grid of {\sc fastwind} models, covering the range of spectral types and luminosity classes, in which we varied the nitrogen abundance relative to hydrogen
\begin{equation}\label{eq:nabun}
\nabun = \log \left(\frac{N_{\rm N}}{N_{\rm H}} \right) + 12,
\end{equation}
where $N_{\rm N}$ and $N_{\rm H}$ are the number abundances of nitrogen and hydrogen. We vary \nabun\ from \nabun\ = 6.93, which is close to the LMC baseline value \citep{kurt1998}, to \nabun\ = 8.53.
Figure~\ref{fig:TMNgrid} shows the outcome of this test for a few representative spectral lines. As expected, most lines only show sensitivity in a given temperature range. On the basis of this exercise we compiled a list of primary and secondary diagnostic lines, which is given in Table~\ref{tab:Nlines}. Primary lines are those most often used for abundance measurements, as they show reliable results. Secondary lines are often not visible or may give anomalous results as explained in the next paragraphs.

\begin{table}[]
\centering
\caption{Diagnostic lines used to derive the nitrogen abundance. See Sect.~\ref{sec:N_spectroscopy} for a discussion of
         the primary and secondary categories of lines. Table~2 of \citet{riverogonzalez2012a} provides the details of specific lines, such as the central wavelength and potential blends.}
\label{tab:Nlines}
\begin{tabular}{ll}
 \hline \hline \\[-9pt]
 \multicolumn{2}{l}{\em Primary diagnostic lines}  \\
 \hline \\[-9pt]
 \ion{N}{ii}    & $\lambda$3995  \\
 \ion{N}{iii}   & $\lambda$4379, 4511$-$4515 \\
 \ion{N}{iv}    & $\lambda$4058 \\
 \ion{N}{v}     & $\lambda$4603$-$4619 \\
 \hline \\[-9pt]
 \multicolumn{2}{l}{\em Secondary diagnostic lines}  \\
 \hline \\[-9pt]
 \ion{N}{ii}    & $\lambda$4447, 4601$-$4607$-$4621$-$4630$-$4643 \\
 \ion{N}{iii}   & $\lambda$4097, 4195, 4518, 4523, 4535, 4634$-$4640$-$4641 \\
 \hline
\end{tabular}
\end{table}

The \ion{N}{iii}\,$\lambda$4634$-$4640$-$4641 triplet in emission, in combination with \ion{He}{ii}\,$\lambda$4686, 
is used to assign O-type stars the qualifier $f$ \citep{sota2011}.  The modelling of this 
triplet is complex and has been described in detail by \citet{riverogonzalez2011}. We confirm 
their findings that the abundances derived from this triplet are sometimes anomalous compared 
to that of other lines and therefore we use it as a secondary diagnostic.

One of the strongest lines, \ion{N}{iii}\,$\lambda$4097, is situated in the wing of H$\delta$. 
To perform an equivalent width analysis, we divided the observed profile by a fit to the H$\delta$ wing as 
if it represented the continuum.
In some cases this method worked, however in other cases anomalous abundances (with respect 
to constraints from other lines) were found. In particular, we found for 15 stars that the \ion{N}{iii}\,$\lambda$4097 profile is underpredicted when adopting the N-value derived from other lines. We thus used this line too as a secondary diagnostic only.

As stars rotate faster and their lines become broader, neighbouring spectral lines can blend with each other. This is in particular the case for the \ion{N}{iii} $\lambda$4511$-$4515$-$4518 complex. In these cases we attempted to extract the combined \Weq\ from the spectrum (see Sect.~\ref{sec:abundance}).
Unfortunately, the signal-to-noise ratio (\SNR) of our data is such that for almost all rapidly rotating ($\vrot\gtrsim170\,\kms$) stars in our sample, the spectral lines could not be distinguished from the noise. 

\citet{riverogonzalez2012a} derived the nitrogen abundances of LMC O-type stars relying mostly on the same lines that we use. \citet{martins2015}, studying Galactic stars using the code {\sc cmfgen} \citep{hillier1998}, rely on 4 to 22 lines that partly overlap with our set of lines. By virtue of their wider spectral range and higher baseline abundance, \citeauthor{martins2015} were able to use more, as well as intrinsically weaker, spectral lines. However, their list does not include our primary lines \ion{N}{iii}\,$\lambda$4379, \ion{N}{iv}\,$\lambda$4058 and  \ion{N}{v}\,$\lambda$4603$-$4619.

It is noteworthy that the only \ion{N}{iv} line available to us ($4058\,\AA$) is also sensitive to effects other than variation of the nitrogen abundance, as described by \citet{riverogonzalez2012b,riverogonzalez2012a}. Unfortunately \ion{N}{iv}\,$\lambda$6380, a spectral line that is known to be strong \citep[e.g.,][]{riverogonzalez2012a}, lies outside of our spectral range.

\subsection{Abundance determination}
\label{sec:abundance}

\begin{figure*}[t!]
\centering
\includegraphics[scale=0.375]{\figpath 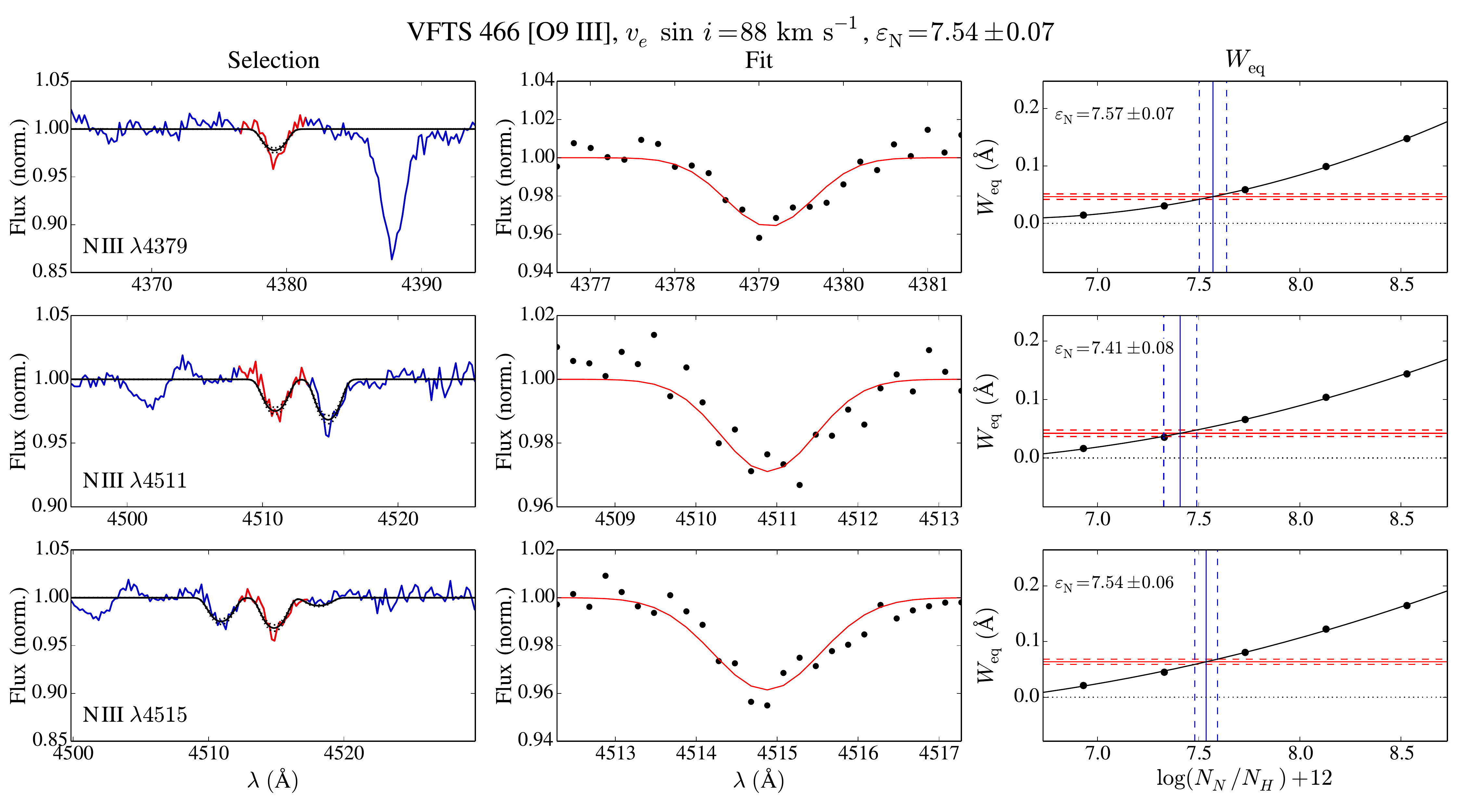}
\caption{Abundance measurement for VFTS\,466, a relatively high S/N source. Not all lines used are
shown. {\em Left panels:} the observed spectrum is shown in blue and the best model
fit in black. The nitrogen abundance of the best model resulted from combining all utilised lines and is given by the \nabun\ in the figure title. The selection of the fit region is displayed in red. 
{\em Middle panels:} zoom in on the fitted region, with the best 
fit Gaussian profile in red. {\em Right panels:} theoretical \Weq\ values as a function of
\nabun\ (horizontal axis) are shown in black. Measured \Weq\ and interpolated \nabun\ are
in red and blue, respectively. Dashed red lines indicate the measurement error and dashed
blue lines the corresponding uncertainty in the best fit value.}
\label{fig:exampledet1}
\end{figure*}
\begin{figure*}[t!]
\centering
\includegraphics[scale=0.375]{\figpath 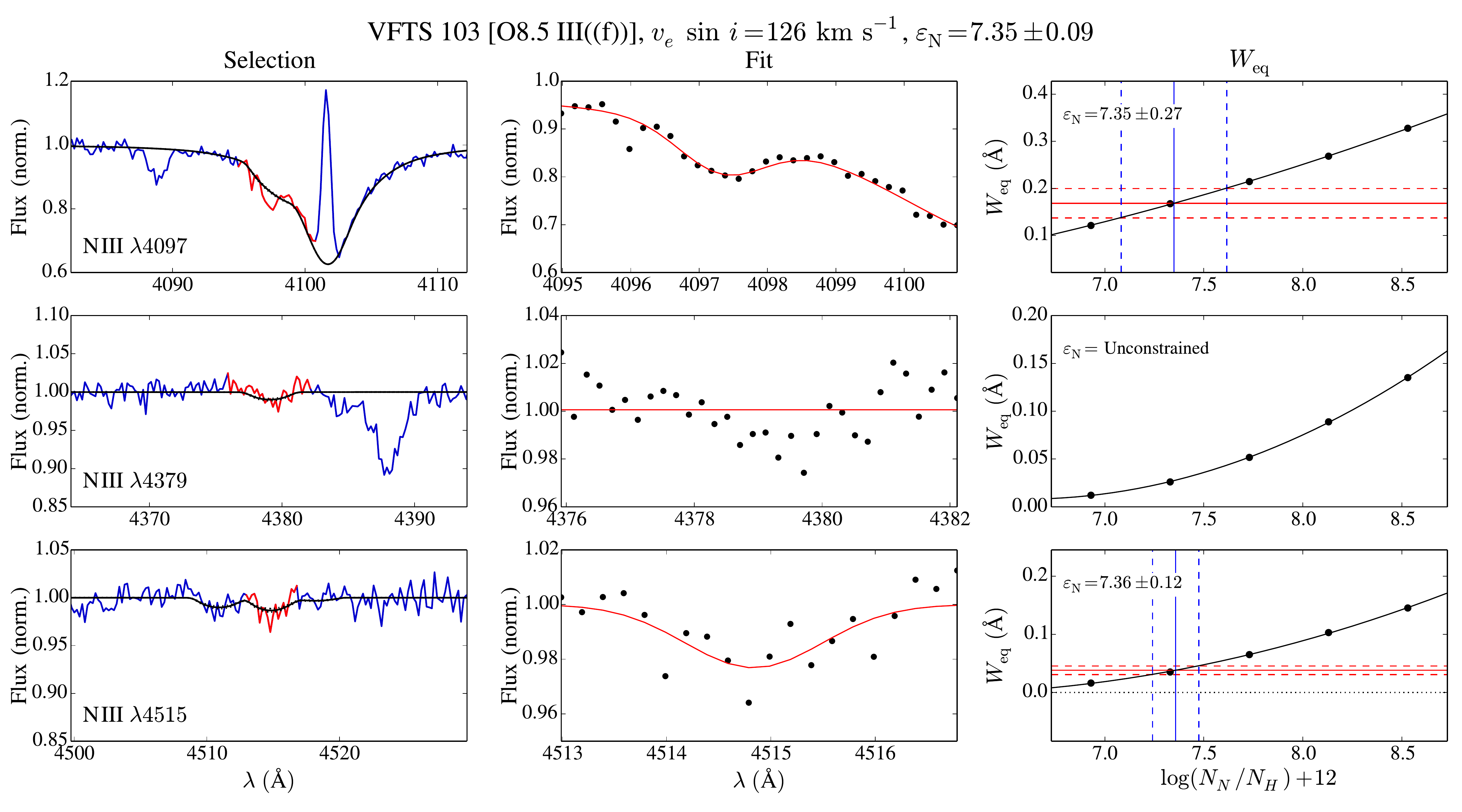}
\caption{Same as Fig.~\ref{fig:exampledet1}, for VFTS\,103, typical for relatively low S/N. In the case of the \ion{N}{iii} $\lambda4379$ line (middle row), no Gaussian profile could be fitted. However, the final line profile that is derived from measurements to other lines, is consistent with the data.}
\label{fig:exampledet2}
\end{figure*}

\begin{figure*}[t!]
\centering

\includegraphics[scale=0.375]{\figpath 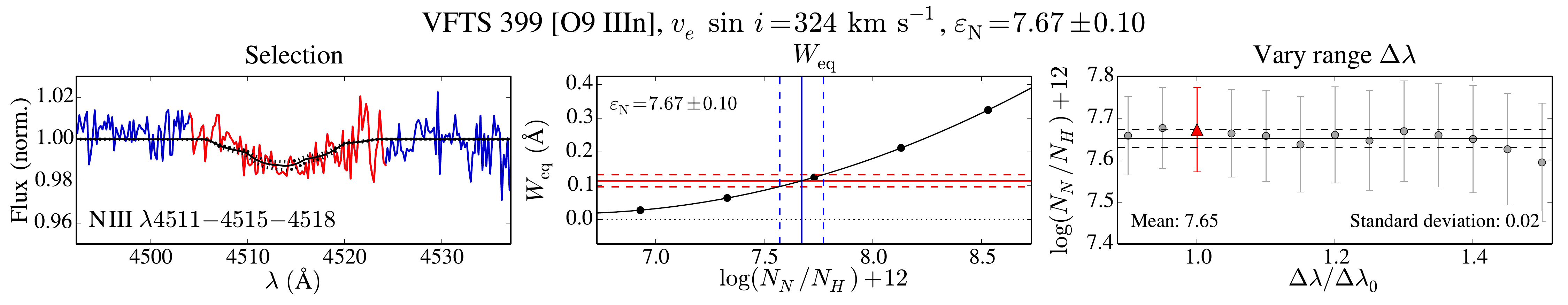}
\caption{ Example of a \Weq\ measurement by integration (for VFTS\,399). The \Weq\ is extracted by integration of the red region of the spectrum in the left panel. In the middle panel, this is compared to theoretical \Weq\ values of the combined complex as predicted by {\sc fastwind}, yielding \nabun\ and its associated error. The robustness of the measurement is checked twofold: First by comparing a model with the final solution to the observed spectrum (left panel, solid black line). Second, in the right panel, by comparing the abundance measurements that result from adjustment of the integration bounds, varying the width $\Delta\lambda$ as function of the adopted width $\Delta\lambda_0$, where the adopted measurement is indicated by the red triangle. The 1$\sigma$ error on \Weq\ is calculated through propagation of the error spectrum.} 
\label{fig:exampleint}
\end{figure*}
\begin{figure*}[th!]\centering
\includegraphics[scale=0.375]{\figpath 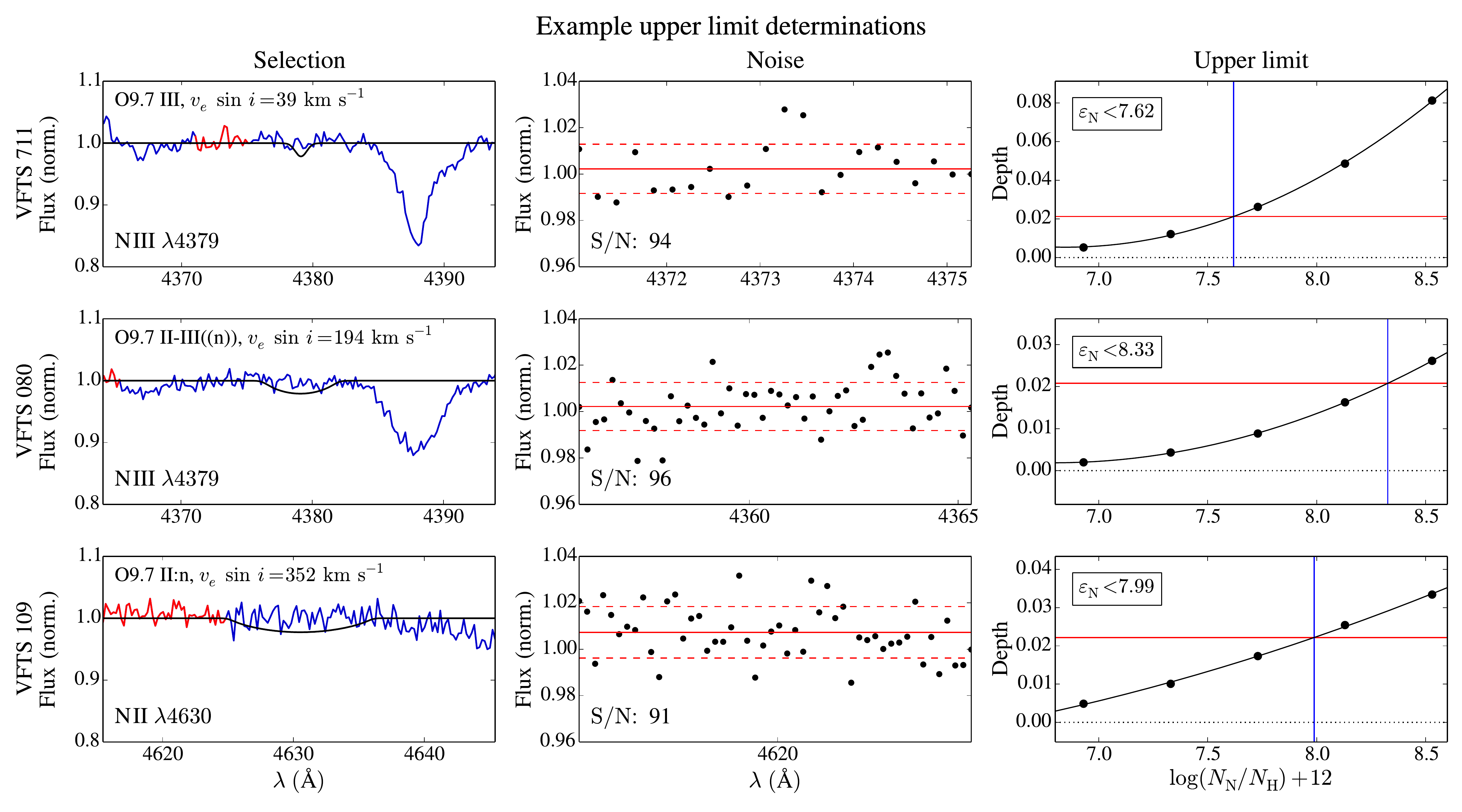}
\caption{VFTS\,711, 080, and 109 are examples of cases for which only upper limits could be derived.
{\em Left panels:} the observed spectrum is shown in blue, the region used to determine the S/N in red, and the model representing the upper
limit in black. 
{\em Middle panels:} zoom in on the fitted region. For these cases this region is used 
to determine the signal to noise. The red line is the mean flux level of this range; 
the dashed lines indicate the standard deviation. 
{\em Right panels:}
theoretical depth values as a function of \nabun\ in black, 
where the minimal detectable depth and its inferred upper limit are indicated as the red horizontal 
and the blue vertical line respectively. The resulting upper limit is indicated in the top left corner of the right panel. The line depression for the upper limit 
to \nabun\ can be considerably larger than the noise would suggest as a result of our conservative choice for this limit (see Sect.~\ref{sec:abundance}).}
\label{fig:exampleulim}
\end{figure*}

We measured the equivalent width \Weq\ of our lines by fitting a Gaussian to their continuum normalised profiles (see the middle columns of Figs.~\ref{fig:exampledet1} and~\ref{fig:exampledet2}). In this procedure, the continuum was held fixed. The centre of the line was allowed to vary within the uncertainty of the RV measurement.  The (weak) nitrogen lines are well represented by a Gaussian profile, the fitting of which provides a straightforward way of extracting \Weq\ and its associated error, given the quality of the data.

Next, using the stellar properties given by Ram\'{i}rez-Agudelo et al. (subm.), we computed a set of five {\sc fastwind} models for each star, in which we varied the nitrogen abundance from the baseline
value \nabun\ = 6.93 to \nabun\ = 8.53 in steps of 0.4 dex. This covers the full range of measured \nabun. The behaviour of the line strength as a function of \nabun\ can be accurately represented on the basis of these points using a second order polynomial, as is shown in the right-most panels of Figs.~\ref{fig:exampledet1} and~\ref{fig:exampledet2}. For each spectral line, the N-abundance and its error (blue solid and dashed lines) was then extracted on the basis of the measured \Weq\ and its error (red solid and dashed lines). 
%
For each star, we combined the results from the different lines that were visible in the spectrum. We computed a weighted mean, where
the N-abundance measured from each line $i$ was weighted according to its error as $1/\sigma_i^2$. 
We considered the weighted standard deviation to be the error on the measurement of \nabun. Independently, the error on the weighted mean can be calculated through propagation of the error on the measurements from individual lines. In cases where the error on the weighted mean was larger than the weighted standard deviation, we adopted the former as error of the measurement. 

These errors do not take into account uncertainties in the determined stellar parameters and in the placement of the continuum. To test the impact of the former on our abundance measurements we computed models ranging over the 2$\sigma$ errors provided by Ram\'{i}rez-Agudelo et al. (subm.). We did this for three stars, representative for late, mid and early spectral type, and for the three parameters that impact our measurement the most, \Teff, $\mathrm{g}$ and $\dot{M}$. For each permutation of the lower and upper errors in the parameters, we repeated our nitrogen abundance measurement. In these measurements, we observed a peak-to-peak spread in \nabun\ of typically $\sim$0.4$\,\mathrm{dex}$, extending up to $0.6\,\mathrm{dex}$ in the case of the hottest star, where the uncertainty in $\dot{M}$ is dominant\footnote{This is because the \ion{N}{iv-v} lines are very sensitive to the adopted $\dot{M}$. In combination with the uncertainty in \Teff, small changes in $\dot{M}$ result in a larger peak-to-peak spread than is the case for the cooler stars, even though $\dot{M}$ is better constrained for hot stars with prominent winds.}. We conclude that the 2$\sigma$ uncertainties in the atmospheric parameters induce an error that may reach up to $\sim$0.2$-$0.3\,dex in extreme cases; hence the 1$\sigma$ uncertainty may reach half of this value. The uncertainty in the location of the continuum  is an additional source of error in the \Weq\ determination. \citet{sana2013} reported that the continuum is constrained to better than 1\% on average. We estimate the corresponding error in the determination of \nabun\ to be up to 0.15\,dex for modestly rotating stars. 
In the course of this work, we will consider the error on our abundance measurements to be the standard deviation of the N-abundance that is measured from different diagnostic lines.

 Figures~\ref{fig:exampledet1} and~\ref{fig:exampledet2} show two examples of the \Weq\ analysis,
 for VFTS\,466 and VFTS\,103, typical for high and low \SNR\ data respectively. In each case,
 results for only a few of the fitted lines are shown. In the right hand columns, the quadratic interpolation of the observed \Weq\ (red horizontal line) with the theoretical \Weq\ (black dots and line) is shown. By computing a model with the final solution, 
 and overplotting it on the data, we test the robustness of the method (left columns of
 Figs.~\ref{fig:exampledet1} and~\ref{fig:exampledet2}).
 
 In some cases a measurement of \Weq\ could not be extracted from the spectrum through fitting a Gaussian profile (e.g., due to blending with neighbouring lines). In these cases we obtained \Weq\ by direct integration of the spectrum. One example, for VFTS\,399, is shown in Fig.~\ref{fig:exampleint}. Keeping the continuum fixed, we integrated the spectral range where the N-lines are predicted to reside (red range in the left panel) and calculated the error on \Weq\ by propagation of the error spectrum. By comparing the observed \Weq\ with the theoretical values from {\sc fastwind}, we extract a \nabun\ measurement and its error (middle panel). By overplotting a model with the final solution on the data (left panel) and comparing the resulting measurements of different integration ranges (right panel), we tested the robustness of the method. This method was employed for the \ion{N}{iii} $\lambda$4511$-$4515$-$4518 complex in 4 cases, namely VFTS\,185, 399, 513, 569. For VFTS\,399, good agreement is found with earlier findings by \citet{clark2015}, who performed a by-eye comparison, confirming the N-enriched nature of the star.
 
 Figure~\ref{fig:examplesenriched} shows additional examples of line fits, for stars across the full range of considered spectral types and N ionisation stages. It shows that, in order to reproduce the N line profiles of these sources, significant surface enrichment is required.

If no lines are detected, an upper limit on \nabun\ can be set. Examples of this are shown in Fig.~\ref{fig:exampleulim}. From the calculated models the strongest spectral lines in terms of depth were selected. For these lines the noise in the continuum was measured and converted into a minimal detectable depth, given by $2\times\sigma_\mathrm{cont}$. In doing so, we assumed the line is detectable when its depth constitutes twice the noise level. The predicted depth as a function of \nabun\ was measured from the models and again described by a quadratic fit. For each line, the minimal detectable depth then yielded the upper limit on \nabun. The resulting upper limit is thus solely based on the (predicted) depth of a line, rather than its \Weq. Assessing the outcome of this method for the strongest 5-7 lines, taking into account their local continuum noise and individual strengths, allowed us to set an upper limit on the nitrogen abundance of a star. We pursued to use the nitrogen line that sets the tightest constraint on \nabun. If we adopted a 1$\sigma$ (instead of the 2$\sigma$) criterion, resulting upper limits are lower by about $0.4\,\mathrm{dex}$.

\subsection{Limitations of the method}
\label{sec:limitations}

\begin{figure}[t!]
\begin{center}
\includegraphics[scale=0.425]{\figpath 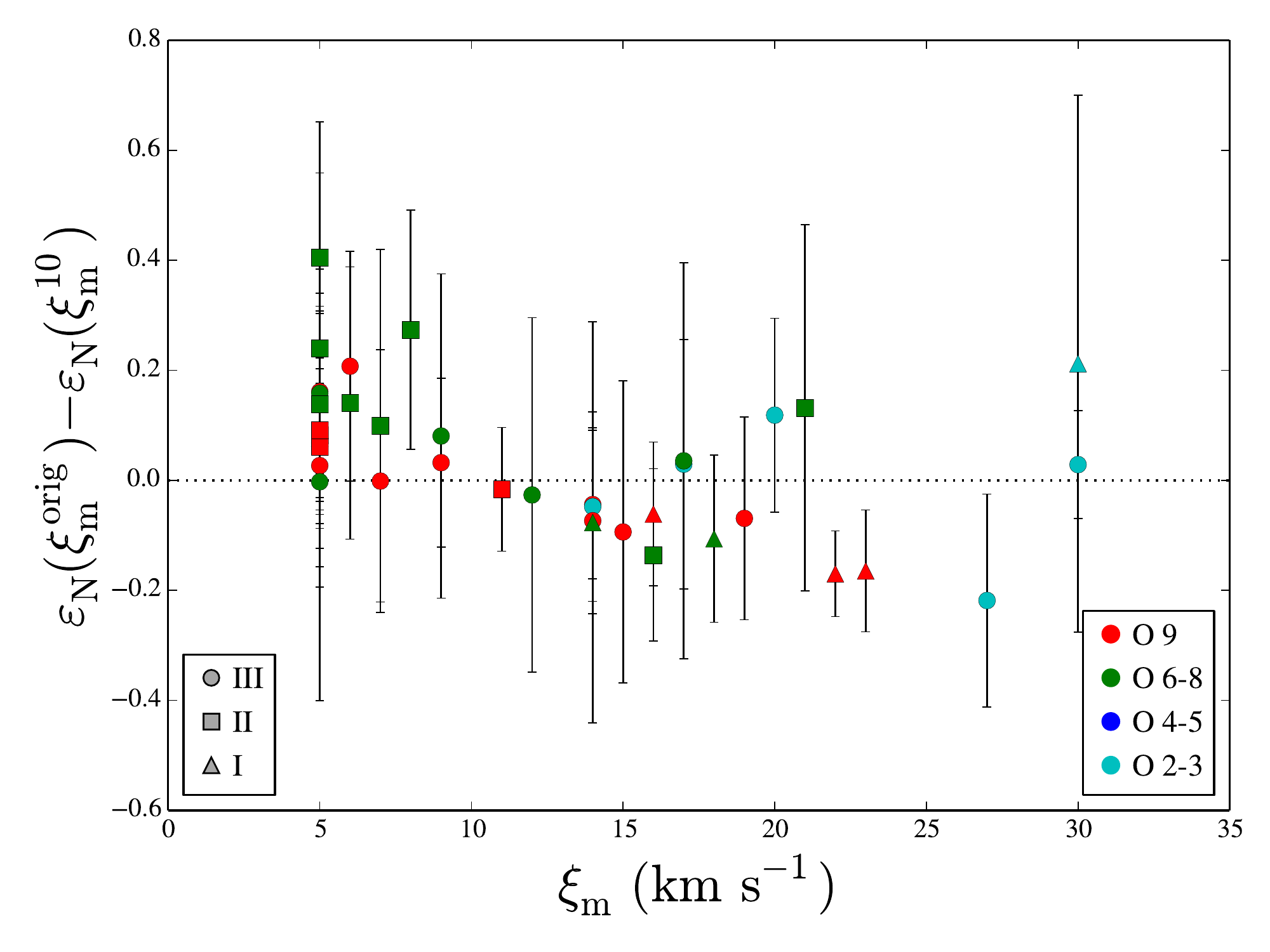}
\caption{The difference in the measured N-abundance using the micro-turbulent
velocity as determined by Ram\'{i}rez-Agudelo et al. (subm.) and $\vmicro = 10\,\kms$.
All stars for which the nitrogen abundance could be determined are shown.  }
\label{fig:vmicro}
\end{center}
\end{figure}

Though the nitrogen lines appear weak, they may suffer from saturation effects and therefore their strength can be a function of the micro-turbulent velocity. The \vmicro\ values, fitted by Ram\'{i}rez-Agudelo et al. (subm.), range between a few and $30\,\kms$, often with sizable uncertainties, and where $30\,\kms$ is the fixed upper boundary considered by these authors. We investigated the effect of the micro-scale turbulent motions on the nitrogen abundance by computing new {\sc fastwind} models with the same atmospheric parameters, but adopting a fixed  $\vmicro$ of $10\,\kms$. Following the method described in Sect.~\ref{sec:abundance} we derived the N-abundance also for this constant value. Since the \Weq\ of a line is in principle a positive function of \vmicro, one expects measurements to yield lower \nabun\ when fitted with a higher \vmicro\ (keeping all other parameters fixed). However, the micro-turbulence may also affect the atmospheric structure and thus the emergent line profiles. In Fig.~\ref{fig:vmicro} we show the difference in \nabun\
between the two methods as a function of the \vmicro\ value from the automated fitting approach.  
The differences are typically less than 0.2 dex. 
 Because the \vmicro\ values are largely unconstrained, i.e. have large uncertainties, by Ram\'{i}rez-Agudelo et al. (subm.), we henceforth adopt $\vmicro=10\,\kms$ in all our computations.


Ram\'{i}rez-Agudelo et al. (subm.) determined the atmospheric parameters without taking into account macro-turbulent broadening of spectral lines. This leads to \vrot\ values that 
in the regime up to $\sim 150\,\kms$ are, on average, somewhat larger than those incorporating macro-turbulent broadening \citep[see][for a discussion]{ramirezagudelo,simondiaz2014,markova2014}. We too neglected macro-turbulent broadening and adopt \vrot\ values given by Ram\'{i}rez-Agudelo et al. (subm.). 
Since rotation and macro-turbulent motions conserve equivalent width, this has a marginal impact on our measurements of the N-abundance.

In the spectra of some rapidly rotating stars, we saw signs of the \ion{N}{iii}\,$\lambda$4097 line. However, due to the severe broadening, we could not extract \Weq\ through a combined Gaussian and Lorentzian fit (the Lorentzian fit was employed to represent the wing of H$\delta$, hence the local continuum of the \ion{N}{iii} line; see Sect.~\ref{sec:N_spectroscopy}). In these cases, spectral synthesis would be the better option. However, we found this line often to be underpredicted in cases where other lines provide independent measurements (see Sect.~\ref{sec:N_spectroscopy}). The \nabun\ required to reproduce the \ion{N}{iii}\,$\lambda$4097 line would imply that other diagnostic lines should be visible. We thus opted in these cases for an upper limit that brings the predicted {\sc fastwind} profiles in agreement with the observed profile of \ion{N}{iii}\,$\lambda$4097, but would yield other visible diagnostic lines.

Our sample contains 6 stars which were, after radial velocity measurements at additional epochs, identified as long-period spectroscopic binaries (Almeida et al. in
prep.). These stars are VFTS\,064, 093, 171, 332, 333, 440. Since their spectra are fitted assuming they were single, their resulting parameters should be considered with caution. As such they are excluded from the quantitative comparison with theory in Sect.~\ref{sec:popsyn_lowg}. VFTS\,399 is excluded in that section as well, since it may be an X-ray binary \citep{clark2015}. The abundance measurements of these 7 stars are flagged in Fig.~\ref{fig:hunterplot_stars}. In the sample of stars with no luminosity class identifier, Ram\'{i}rez-Agudelo et al. (subm.) find no satisfactory fit to the spectral lines of another 6 objects. Finally, we are unable to reliably extract the nitrogen abundance of one source (VFTS~177, see also Appendix~\ref{app:datatablenoLC}). These further 7 objects are also excluded from the analysis in Sect.~\ref{sec:popsyn_lowg}.


\section{Results}
\label{sec:results}



\begin{figure}[th!]
\centering
\def\hunterplotscale{0.42}
\includegraphics[scale=\hunterplotscale]{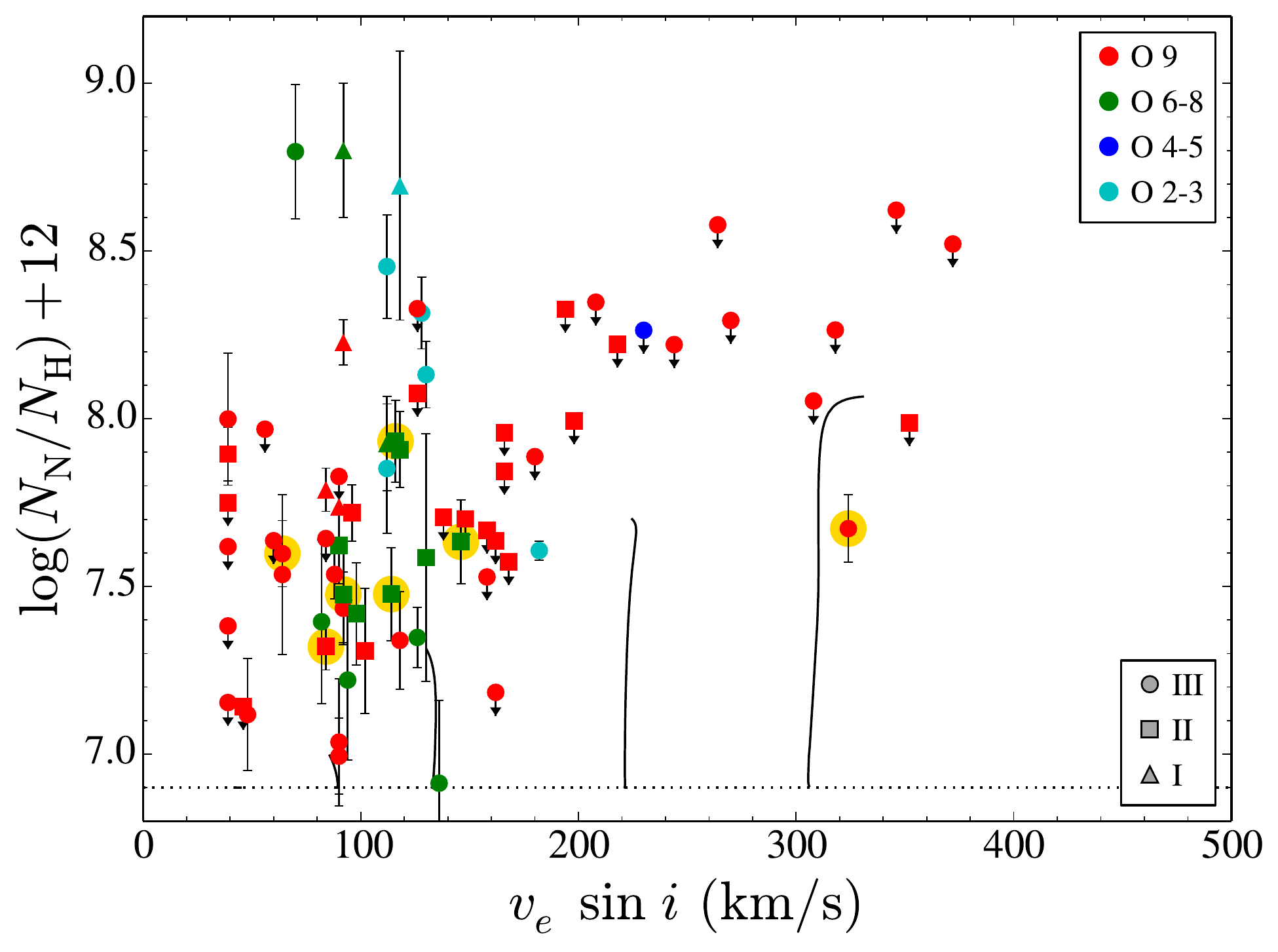}
\includegraphics[scale=\hunterplotscale]{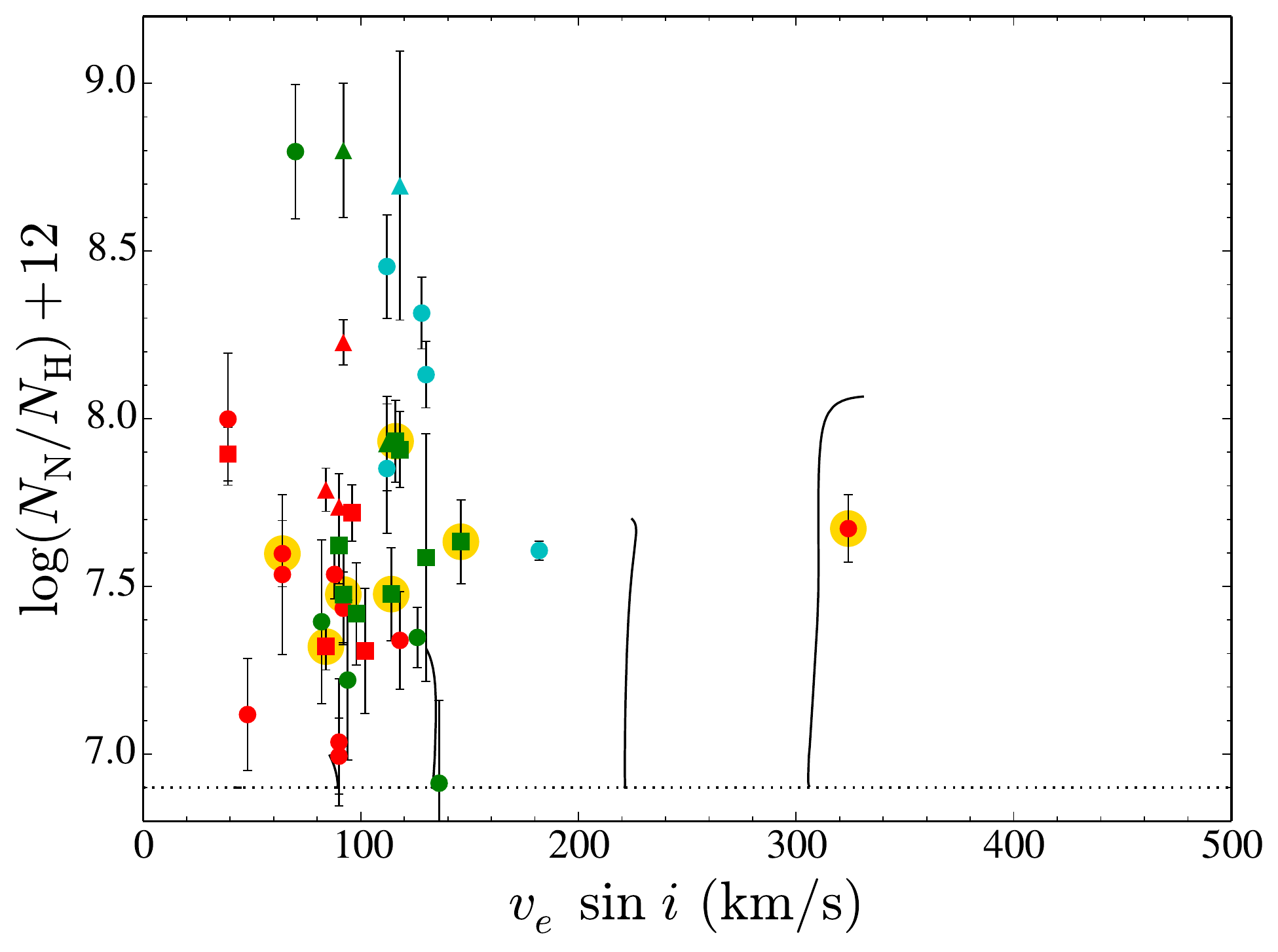}
\includegraphics[scale=\hunterplotscale]{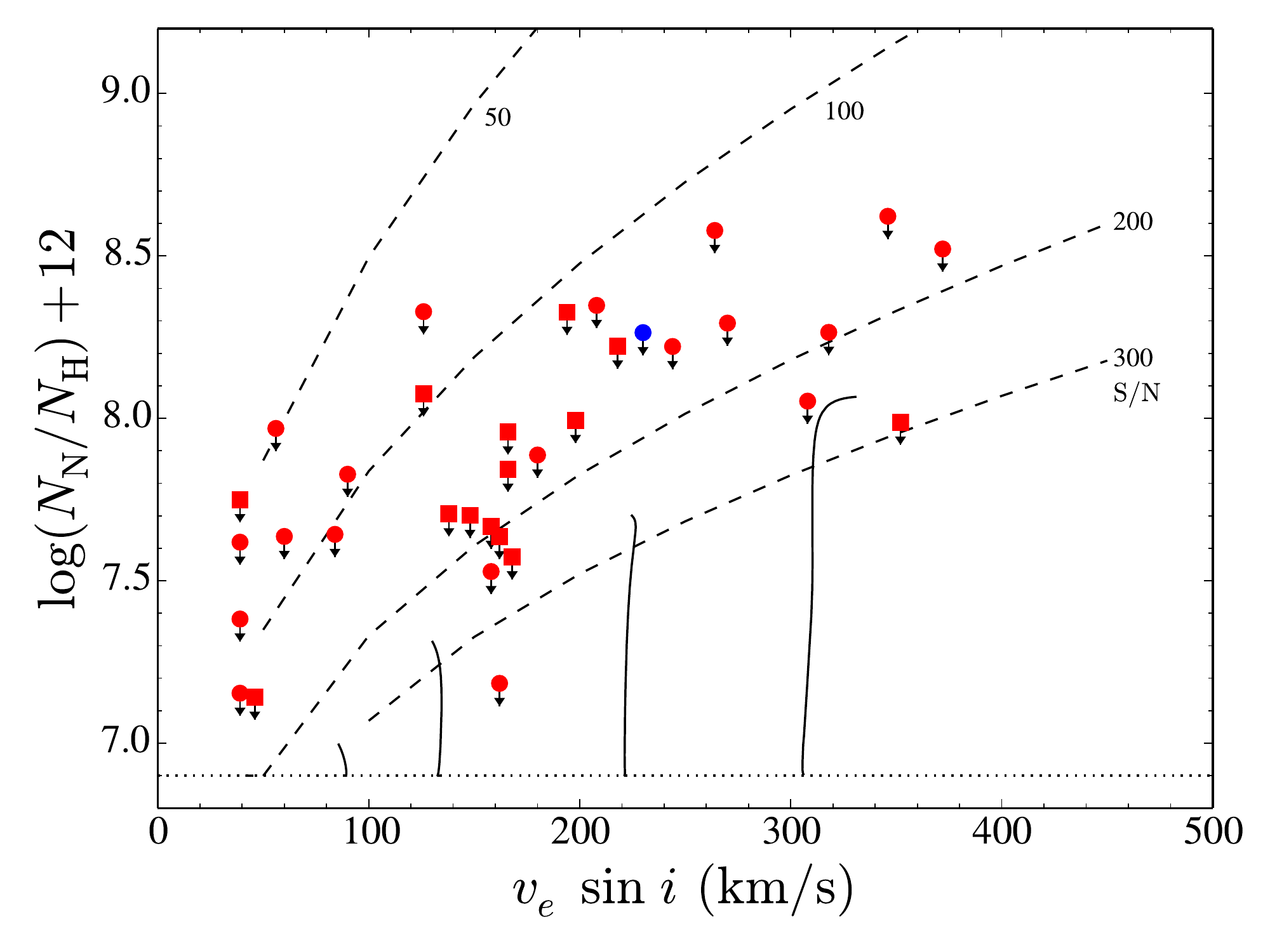}
\caption{Nitrogen abundance versus projected rotational velocity. For all panels, symbols 
denote luminosity class and colours spectral type; see legend in the top panel. 
{\em Top panel:} Entire sample. The newly detected binaries have a yellow circle underlaid (see Sect.~\ref{sec:limitations}). {\em Middle panel:} The subset of 
stars for which the nitrogen
abundance could be measured.  {\em Bottom panel:} The subset of stars
for which only upper limits could be derived. The dashed lines indicate detection limits
for different \SNR\ and are discussed in Sect.~\ref{sec:results}. The black solid lines are evolutionary tracks for the O-type stage ($\Teff\geq29\,\mathrm{kK}$) of a 20\,\msun\ star \citep{brott2011a}, with initial rotational velocities of 114, 170, 282 and 390$\,\kms$. We multiply the rotational velocity of these tracks by $\pi/4$ to account for the average projection effect.}
\label{fig:hunterplot_stars}
\end{figure}
\begin{table}[b!]
\caption{Number of stars for which the nitrogen abundance could be constrained and
for which only upper limits could be derived, as a function of luminosity class.}
\label{tab:sampleparams} 
\centering 
\begin{tabular}{r ccc c}
\hline\hline\\[-9pt]
LC & III & II & I & Total \\
\hline\\[-9pt] 
Abundance constraint &  20 & 12 &  6 &  38 \\
Upper limit &  20 & 14 &  0 &  34 \\
\hline\\[-9pt] 
Total &  40 & 26 &  6 &  72 \\
\hline 
\end{tabular}
\end{table}


The derived surface nitrogen abundances and their uncertainties are listed in
Table~\ref{tab:nresults} of the appendix. In Fig.~\ref{fig:hunterplot_stars} they  
are plotted as function of the projected rotational velocity. 
Note that these values of \vrot\ are, strictly, upper limits, due to the neglect of macro-turbulent broadening.
The symbol type denotes the luminosity class. Of the 72 sources studied, we constrained 
\nabun\ in 38 cases and had to settle for upper limits in 34 cases; see also Table~\ref{tab:sampleparams}. The 7 newly detected binaries in the sample are identified by a yellow circle (see Sect.~\ref{sec:limitations}).

Also shown are evolutionary sequences by \citet{brott2011a} for a $20\,\msun$ star during its O-type phase ($\Teff\geq29\,\mathrm{kK}$). The surface rotational velocity of the tracks from \citeauthor{brott2011a} is essentially constant during its main-sequence evolution. This is the result of a competition between angular momentum transport from the core to the envelope, and envelope expansion. 

Except for VFTS\,399, the sources for which the nitrogen abundance could be determined have \vrot\ up
to 170 \kms. They span a fairly wide range of \nabun\ values, from about 7.0 to 8.8, and
are shown separately in the middle panel of Fig.~\ref{fig:hunterplot_stars}.
Stars with $\nabun \gtrsim 8.1$ are predominantly early-type stars. Also, the
supergiants are all quite enriched having $\nabun \gtrsim 7.7$. 

For all but one star spinning faster than 170 \kms\ we could only derive upper limits.
These are essentially set by the signal-to-noise ratio  of the spectrum and \vrot. 
This is illustrated in the bottom panel of Fig.~\ref{fig:hunterplot_stars}. The dashed lines 
show detection limits covering the range of \SNR\ of our sample stars. They have been computed 
for a star of spectral type O9, representing the bulk of our sample, by considering the primary
diagnostic line \ion{N}{iii}\,$\lambda$4379. For each \SNR\ we established the minimum
detectable depth from the inverse of the \SNR, for a range
of \vrot. For each \vrot, we compared the minimal detectable depth to model spectra for
a range of \nabun\ values (at the appropriate \vrot). This procedure yielded the
detection limits drawn in the panel. 
Given that the typical \SNR\ of our data is $\sim$100$-$200, one may indeed have 
anticipated difficulties in constraining the nitrogen abundance at $\vrot > 200\,\kms$.


\section{Comparison to population synthesis}
\label{sec:popsyn}

In this section we compare our findings to the nitrogen enrichment pattern that is expected for
a comparable (synthetic) population of single O-type stars. This allows us to test models of stellar
evolution in the initial mass range of 15\,\msun\ and up, specifically the treatment of
rotation induced mixing. We limited this test to models by \citet{brott2011a},
as these cover a wide range of initial spin rates and have initial CNO abundances that
are tailored to the LMC environment.


\subsection{Population synthesis model}
\label{sec:popsyn_full}

\begin{figure}[h!]
\centering
\resizebox{1.05\hsize}{!}{\includegraphics{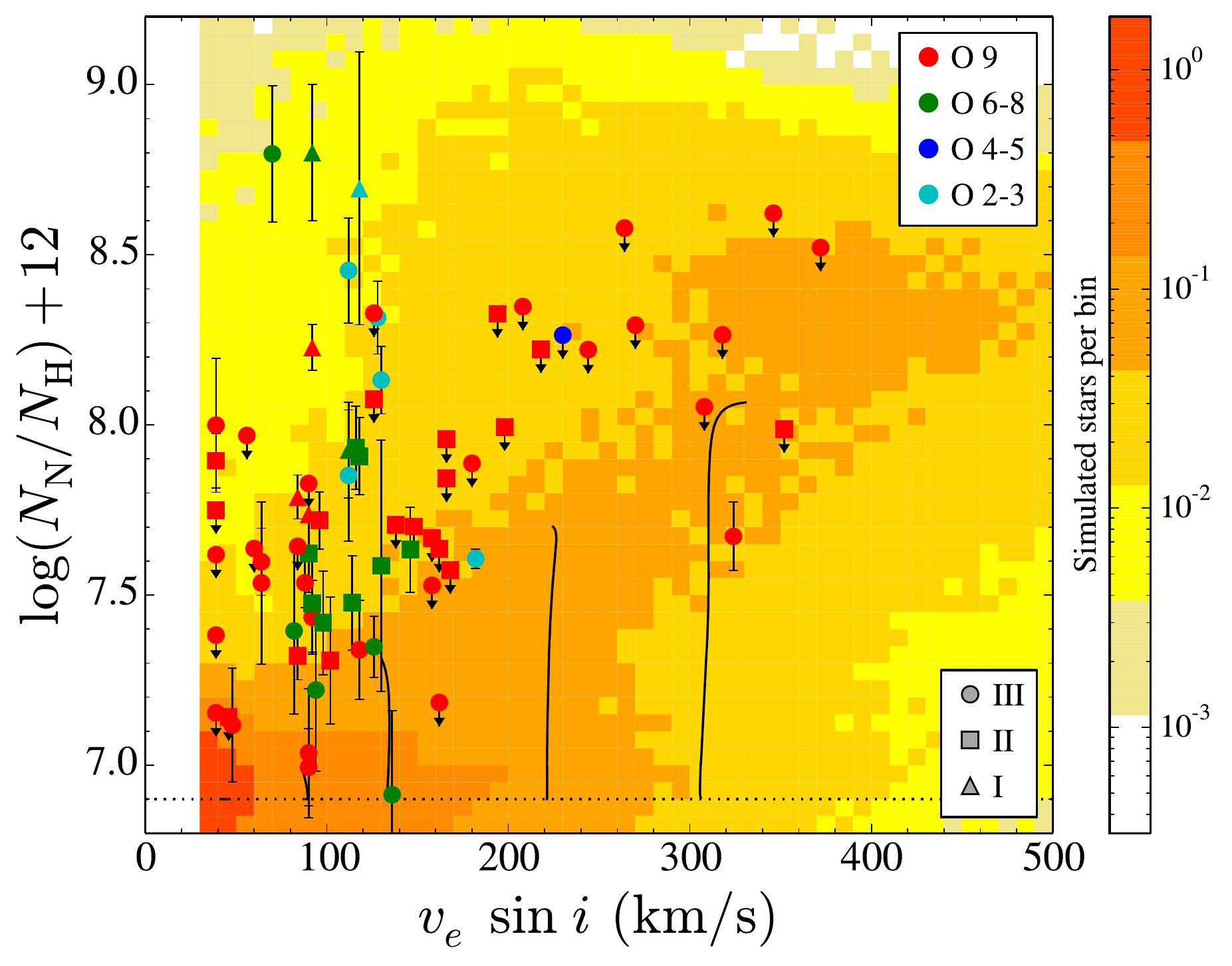}}
\caption{Nitrogen abundance versus projected rotational velocity for the full sample. Symbols indicate luminosity class, colours indicate spectral subtype. 
Tracks \citep{brott2011a} drawn in solid black lines are for a 
$20\,\msun$ star, evolved until the effective temperature is below 29\,kK and they 
become B-type stars. Initial rotational velocities for the tracks are 114, 170, 282 and 390\,\kms\
and have been multiplied by $\pi/4$ to account for the average projection effect. The 
population synthesis calculation projected in the background is for the O star phase only and 
assumes a flat distribution of rotational velocities. Out of $740\cdot10^3$ simulated stars, $310\cdot10^3$ pass the selection criterion ($\Teff \geq 29\,\mathrm{kK}$). These are normalised to 72 stars, matching the 
number of observed stars in the plot.}
\label{fig:hunterplot}
\end{figure}

Figure~\ref{fig:hunterplot} again shows the projected spin velocity versus nitrogen abundance
for all stars in our sample, with symbol shapes and colours as in Fig.~\ref{fig:hunterplot_stars}. In 
the background we show the results of a population synthesis computation for O-type stars, based on 
models by \citet{brott2011a}. These computations were done in the following way.
%
We randomly drew a population of single stars from the evolutionary sequences of \citet{brott2011a},
sampling distribution functions for their initial mass, spin rate and spin axis orientation, and 
adopting continuous star formation for $15\,\mathrm{Myr}$. 
Masses follow a \citet{salpeter1955} initial mass function (IMF)
from 10\,\msun\ to 60\,\msun. After sampling the IMF, the selected mass was rounded off to the mass
of the nearest available evolutionary track.\footnote{This is a simplified approach compared to \citet{brott2011b}. To test its validity, we simulated a population of stars under the same assumptions as \citeauthor{brott2011b} We found small differences that do not impact the conclusions drawn in this paper.} The initial rotational velocity and age were interpolated between tracks. We assumed the spin axes to be randomly distributed in 3D space.

The initial rotational velocity distribution used to construct Fig.~\ref{fig:hunterplot}
is flat and ranges from 0 \kms\ to 600 \kms, to clearly illustrate the effect of rotational
mixing as a function of the projected spin rate. In later figures, we will use the empirical rotational velocity distribution for single O-type stars
derived by \citet{ramirezagudelo} as input.

By accepting a drawn star only if its effective temperature
is at least 29\,kK, we constructed a probability distribution for O-type stars only.
After the star was drawn, we imposed Gaussian distributed uncertainties of 0.15 dex in \nabun\ and 20\% in the rotational 
velocity, characteristic of the typical empirical errors. Drawn stars with $\vrot<40\,\kms$ were artificially placed at $40\,\kms$, as their spin rates are below the resolution limit of the survey. The results were then grouped 
in 10\,\kms\, $\times$ 0.05\,dex bins to generate the plot. Finally, the outcome of the simulation
was normalised such that the simulated star count integrated over the plot equals the number of observed stars 
in the figure. Due to the measurement errors imposed on the simulated values, occasionally stars will acquire a nitrogen abundance smaller than $6.8$. As the minimum N-abundance that we display in the plots is 6.8, i.e., 0.1 dex below
the adopted initial nitrogen abundance for the LMC, we decided to redistribute simulated 
points with \nabun\ $<$ 6.8 to the range [6.8, 7.0] to also visually display the proper normalisation.

The overall predicted trend in Fig.~\ref{fig:hunterplot} is one of increasing \nabun\ with 
increasing \vrot. The reason for the relative dearth of nitrogen normal rapidly rotating stars is
that the rate of nitrogen enrichment as a function of the fraction of the main-sequence lifetime
is a rather strong function of initial mass: the higher $M_{\rm init}$ the faster the
nitrogen enrichment occurs \citep{brott2011b,kohler2012}. For example, a 10\,\msun\ star having
an initial rotational velocity of $\sim$225 \kms\ reaches halfway along its main-sequence a nitrogen
enrichment that is $\sim 40$\% of its final main-sequence enrichment. For a 30\,\msun\ star
halfway along its main-sequence evolution this is already $\sim 85$\%.
This implies that for early-B type stars the lower right corner of the diagram is expected
to be more frequently populated than for O-type stars.

\subsection{Population synthesis for the O\,III-I sample}
\label{sec:popsyn_lowg}

Figure~\ref{fig:hunterplot} does not allow for a direct comparison between our observations and theory because the adopted (flat) spin distribution is not the observed one, and our sample of O-type stars is for giants to supergiants only, i.e., it lacks O-dwarfs. To facilitate a quantitative assessment, we
limited ourselves to the stars that have surface gravities corrected for rotation of $\logg_c \leq 3.75$, for observed as well as simulated stars \citep[see][for a discussion on the centrifugal correction]{repolust2004}. The criterion was chosen as a balance between including as many O giants as possible, while excluding as many O dwarfs as possible. In pursuit of completeness, we supplement 
these O giants
with presumed single O stars that have no luminosity class identifier (see App.~\ref{app:datatablenoLC}). 

In the VFTS sample of presumed single O-dwarfs, analysed by \citet{sabin_sanjulian2014}, only three sources have surface gravities below this limit. These are VFTS\,419, 484 and 581. Preliminary analysis (S\'{i}mon-D\'{i}az, private communication) shows that these stars spin relatively slowly, 145, 120 and 70\,\kms, respectively, and show modest surface enrichments, \nabun\ $\sim$7.2. These three dwarfs are not included in the following sections. We also do not include the seven binaries and seven sources with no luminosity class identifier that are rated as poor quality fit (see Sect.~\ref{sec:limitations}).

\begin{figure}[t!]
\resizebox{\hsize}{!}{\includegraphics{\figpath 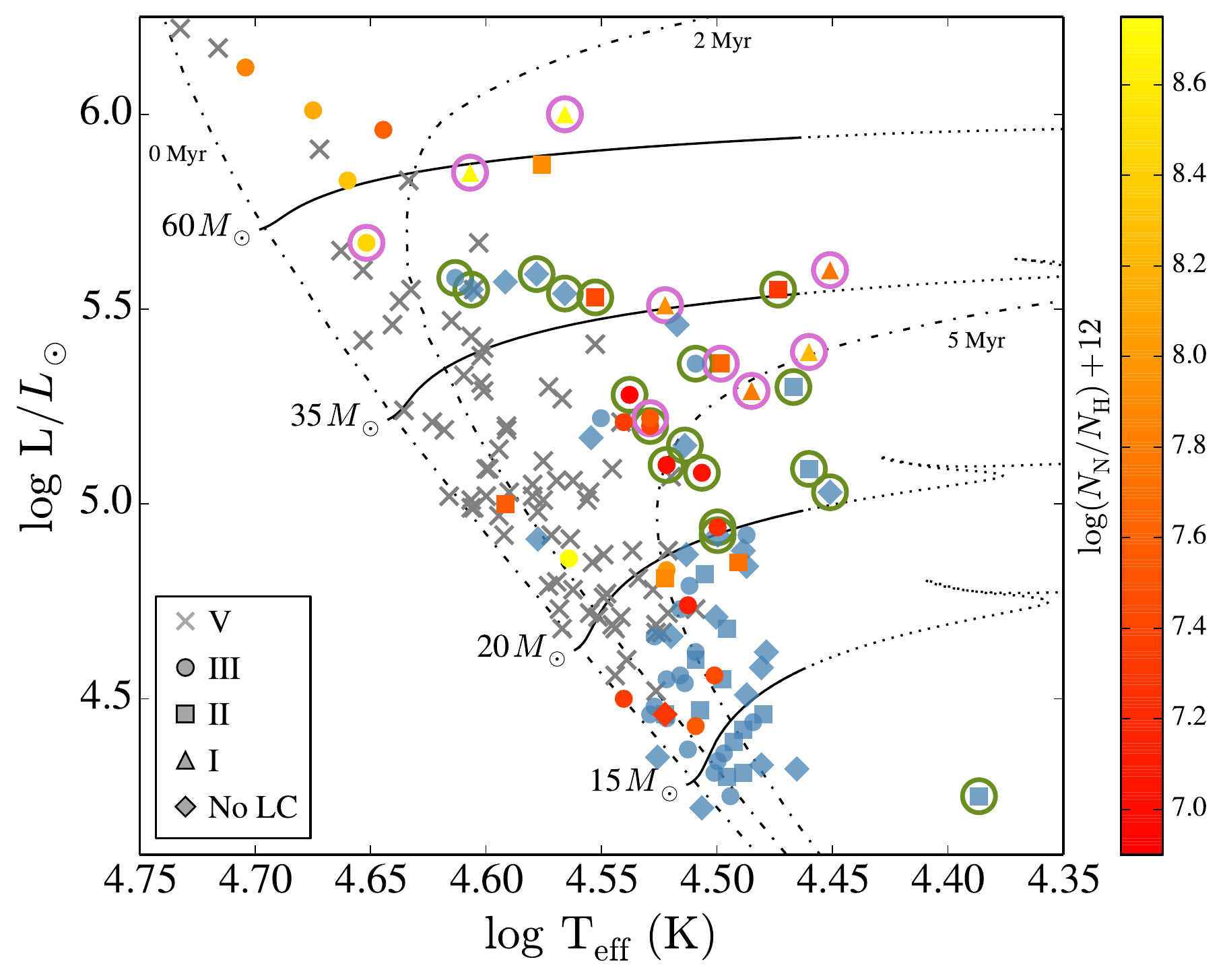}}
\caption{Hertzsprung-Russell Diagram (HRD) for the O-type stars in the VFTS sample, excluding binaries and poor quality fits (Sect.~\ref{sec:limitations}). Luminosities and temperatures are from Ram\'{i}rez-Agudelo et al. (subm.). Evolutionary sequences and isochrones are from \citet{brott2011a} and \citet{kohler2015}. The solid tracks alter to dotted tracks for $\Teff \leq 29\,\mathrm{kK}$. Grey crosses indicate LC\,{\sc V} VFTS stars \citep{sabin_sanjulian2014}. The remaining 
stars are colour-coded according to \nabun, where upper limits are indicated in blue. Encircled stars have $\logg_c \leq 3.75$, where pink circles indicate that they occupy Box~2 and green circles that they are outside of that box (see Sect.~\ref{sec:popsyn_lowg} and Fig.~\ref{fig:hunterplotgroups}).  } 
\label{fig:HRD}
\end{figure}

The total sample of stars with $\logg_c \leq 3.75$ comprises 27 stars (21 O\,{\sc III-I} and 6 without an assigned luminosity class). The restriction to lower gravities overall selects the somewhat evolved stars, 
for which our sample is near to complete within the VFTS. Figure~\ref{fig:HRD} illustrates
this by showing the location of these low gravity sources in the HRD relative to other O-type stars in 
the VFTS, using a colour coding to indicate their N-abundance. 
The gravity criterion also excludes the late O stars ($M\approx 15\,\msun$) that present an ambiguity between the two luminosity class criteria discriminating giants from dwarfs (see \citeauthor{walborn2014} \citeyear{walborn2014}; Ram\'{i}rez-Agudelo et al. subm., for a discussion).

\begin{figure*}[th!]
\centering
\resizebox{0.7\hsize}{!}{\includegraphics{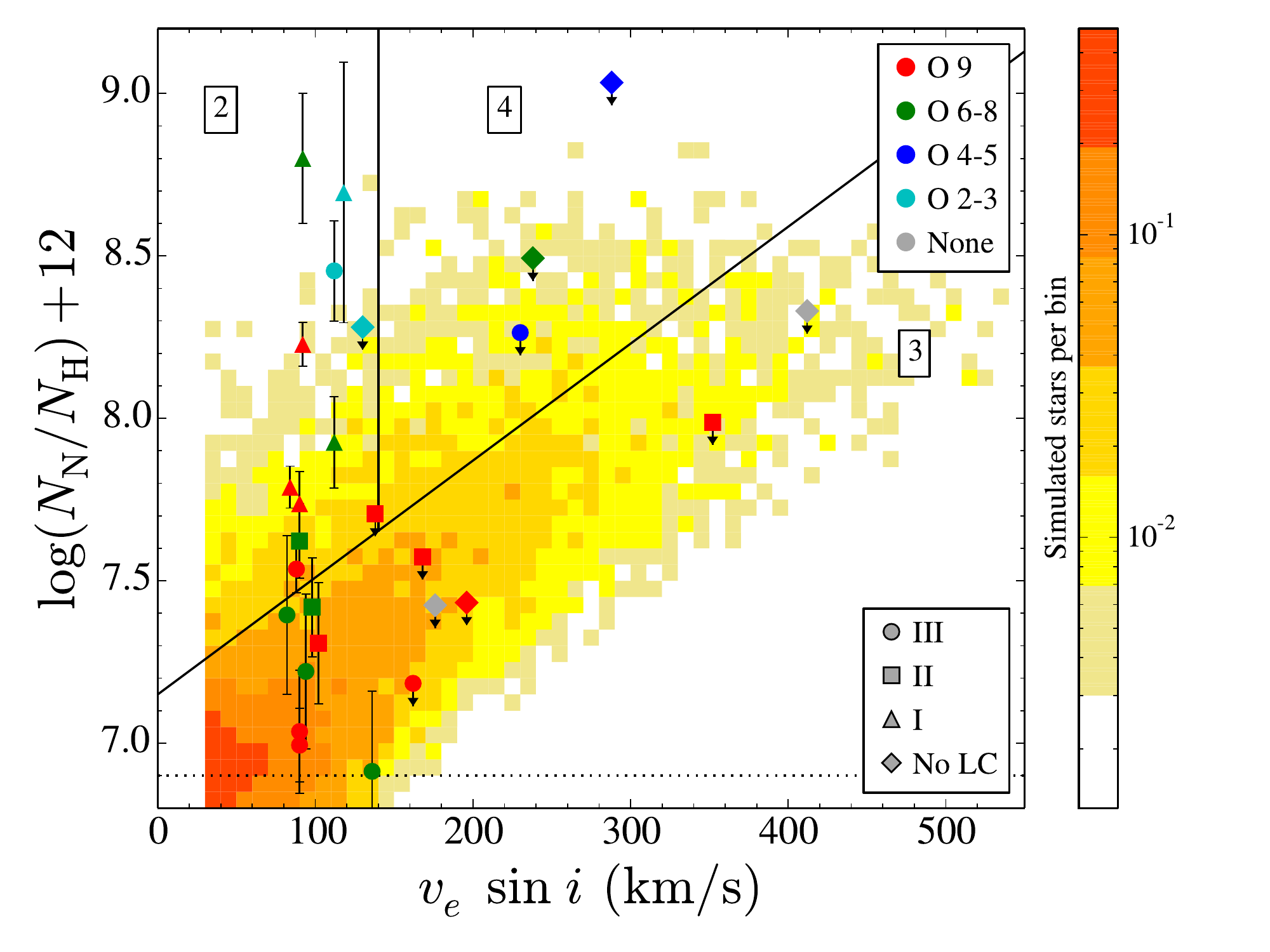}}
\caption{Nitrogen abundance versus projected rotational velocity for 27 O-type stars with gravities 
$\logg_c \leq 3.75$.
It constitutes all of the VFTS sources fulfilling these criteria, save for three
O-dwarfs studied by \citet{sabin_sanjulian2014} that would all end up in Box~3 (S\'{i}mon-D\'{i}az,
private communication), binaries and poor quality fits (Sect.~\ref{sec:limitations}). Identical selection criteria have been applied to the population synthesis
calculation projected on the background. The applied initial spin distribution in this calculation
is assumed to be that derived for the presumed-single VFTS stars \citep{ramirezagudelo}. Out of the $721\cdot10^3$ simulated stars, $21\cdot10^3$ pass the selection criteria ($\Teff \geq 29\,\mathrm{kK} $ and $ \logg_c \leq 3.75$).
These are normalised to 27 stars, matching the number of observed stars in the plot.
The boxes, separated by the solid black lines, are defined in Sect.~\ref{sec:popsyn_lowg}.}
\label{fig:hunterplotgroups}
\end{figure*}

Figure~\ref{fig:hunterplotgroups} shows the subset of low gravity sources (excluding the three O dwarfs)
together with a population synthesis prediction suitable for direct comparison, i.e., for stars of which
the initial spin distribution is as given by \citet{ramirezagudelo}, that have temperature $\Teff \geq 29$\,kK 
and gravity $\logg \leq 3.75$. The population synthesis calculation 
is normalised as to produce the number of observed stars (27) if integrated over the predicted range of \nabun\ and spin rates.

Following \citet{brott2011b}, in Fig.~\ref{fig:hunterplotgroups} we define boxes in which we count the number of 
observed and predicted stars. For the sake of clarity, the numbering of these boxes is chosen such
as to best correspond to the box-definitions used in that paper:

\begin{itemize}
\item[\em Box~2$\!\!\!\!\!\!\!\!\!$] $\,\,\,\,\,\,\,\,\,$selects the N-enriched 
stars with $\vrot < 140\,\kms$.
\item[\em Box~3$\!\!\!\!\!\!\!\!\!$] $\,\,\,\,\,\,\,\,\,$comprises the zone where the effects of rotational mixing are expected to be the most
pronounced. Stars in this region in principle behave in conformity with theory if the bulk is positioned
on or near the predicted coloured diagonal. 
\item[\em Box~4$\!\!\!\!\!\!\!\!\!$] $\,\,\,\,\,\,\,\,\,$is a region where neither stars are observed, nor a significant number of them is predicted.
\end{itemize}
The diagonal separating Boxes 2 and 4 from Box 3 is described by
\nabun\ = $0.0036\,\vrot\ + 7.15$. This division is chosen to follow the dark orange diagonal in Figs.~\ref{fig:hunterplot} and \ref{fig:hunterplotgroups}, i.e., to encompass the bulk of the simulated rotationally-mixed stars. \citet{brott2011b} define Box~1 to select non-enriched rapidly-rotating stars (the lower right of Fig.~\ref{fig:hunterplotgroups}), for which our data do not permit firm conclusions. 

Table~\ref{tab:countstars} shows the counts of the observed and simulated stars in the different boxes. 
In the process of counting we took a Gaussian distributed uncertainty in the observed nitrogen 
abundance into account. 
Upper limits were uniformly distributed between boxes, considering the distance between their value and $\nabun=6.8$. This can result in a non-integer number of stars occupying a certain box in Table~\ref{tab:countstars}.

\begin{table}[t!]
\caption{Star counts and percentages of the total of observed and simulated stars in the boxes 
defined in Fig.~\ref{fig:hunterplotgroups}. 
The $1\sigma$-errors quoted are computed according to binomial statistics.}
\label{tab:countstars} 
\centering 
\begin{tabular}{r rr rr r}
\hline\hline\\[-9pt] 
Box & \multicolumn{2}{c}{Observed} & \multicolumn{2}{c}{Simulated} & Obs/Sim \\
    & \# & \%                      & \# & \% &  \\
\hline\\[-9pt] 
2 &  $10.2 \pm  2.5$ & $37.8 \pm  9.3$ &     2.1 &   7.6  &  4.9 \\
3 &  $15.9 \pm  2.6$ & $59.0 \pm  9.5$ &    22.8 &  84.3  &  0.7 \\
4 &  $ 0.9 \pm  0.9$ & $ 3.2 \pm  3.4$ &     2.2 &   8.1  &  0.4 \\
\hline 
Total &  27 & &  27 & & \\
\hline 
\end{tabular}
\end{table}

We found 60$\pm$10\% of the sample to be in Box~3, containing relatively slowly spinning stars that have 
at most marginally enhanced nitrogen abundances 
and stars for which only upper limits to \nabun\
could be constrained. Theory predicts 85\% of the stars to be in this regime, which appears in reasonable
agreement. However, if future measurements with higher \SNR\ data show that all upper limits in this box
were close to the LMC baseline abundance, this would be
inconsistent with theoretical predictions.
As we cannot investigate this with the data at hand, we conclude that this population is not in 
violation with the predictions of rotational mixing. 

Box~2 contains a population of apparently slowly spinning but nitrogen enhanced stars
(in Fig.~\ref{fig:HRD} these sources have been given a pink circle around their symbol).
While theory predicts 8\% of the stars to be in this region, we found 38$\pm$9\% of the population to be in 
this box. This amounts to almost 5 times more observed sources than predicted sources. It is  
unlikely that this large overabundance of observed sources can be explained by statistical fluctuations 
in the spin axis orientation distribution of these stars, assumed to be random in 3D space. This 
statement gains more weight by the fact that Box~4 is essentially empty, while population synthesis 
predicts an equal amount of sources to reside in this region as in Box~2. We conclude that the bulk
of the sources in Box~2 are truly slow rotators and that their high nitrogen abundance
is not explained by the evolutionary models of \citet{brott2011a} (see also Sect.~\ref{sec:bonnsaicomp}).

\section{Discussion}
\label{sec:discussion}
In this section we discuss the nature of the stars in Box\,2 that do not concur with
predictions of rotational mixing in massive stars. We also compare our finding to previous 
studies. We start by examining the current mass of the stars in our sample.

\subsection{Current evolutionary masses}
\label{sec:massdependence}


Current evolutionary masses of the stars in our sample have been determined by
Ram\'{i}rez-Agudelo et al. (subm.) using {\sc Bonnsai}\footnote{The {\sc Bonnsai} 
web-service is available at \newline \url{www.astro.uni-bonn.de/stars/bonnsai}.}, a Bayesian method 
to constrain the evolutionary state of stars \citep{schneider2014}. As independent prior
functions these authors adopt a \citet{salpeter1955} initial mass function, an initial
rotational velocity distribution as given by \citet{ramirezagudelo}, a random orientiation
of spin axes, and a uniform age distribution. The observables on which the mass estimates
are based are $L$, \Teff\ and \vrot.

\begin{figure}[t!]
\begin{center}
\resizebox{0.8\hsize}{!}{\includegraphics{\figpath 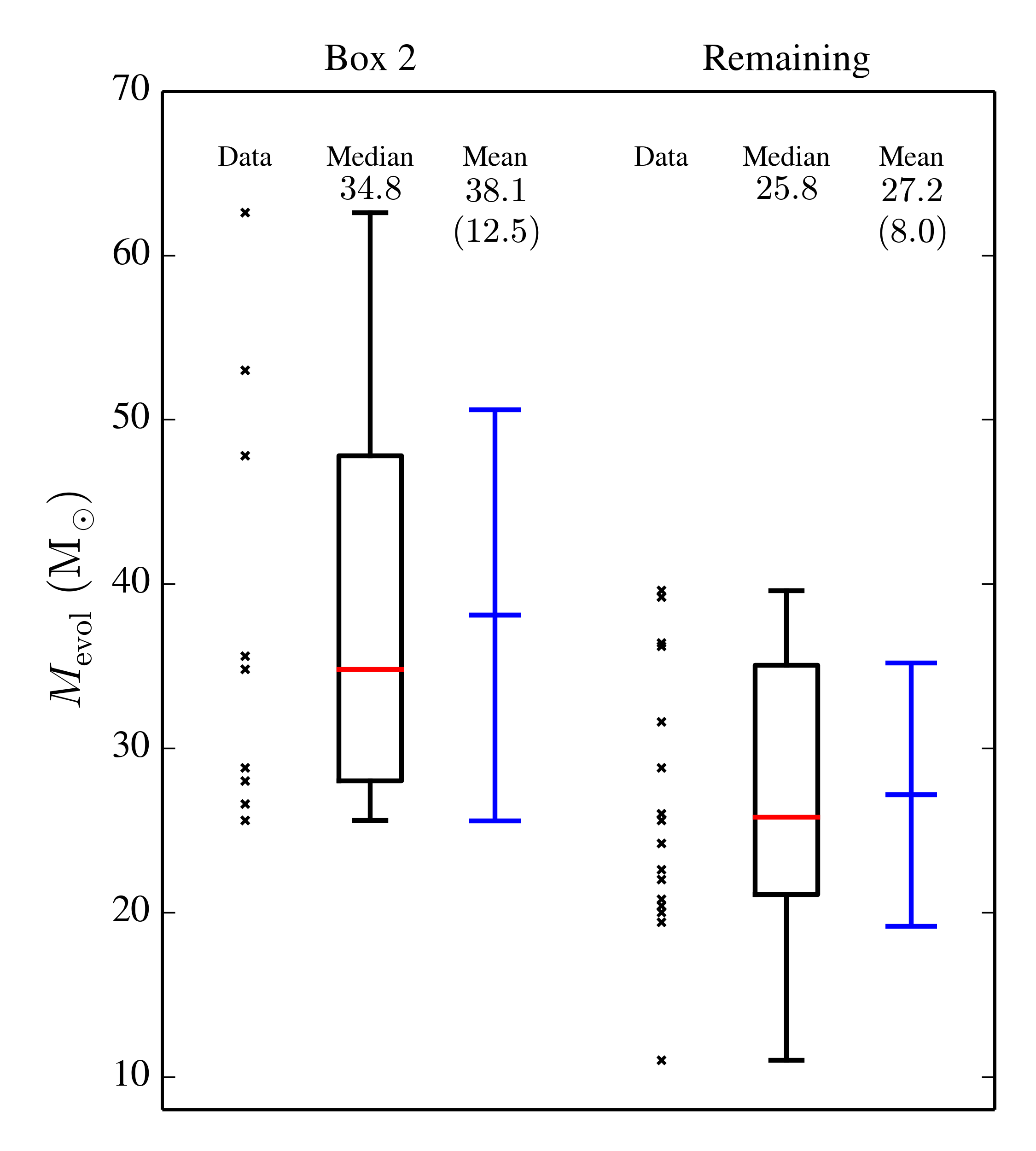}}
\end{center}
\caption{Distribution of evolutionary masses for the stars in Fig.~\ref{fig:hunterplotgroups} that occupy Box~2 (left) and the remainder of stars (right). Black crosses indicate data points. The middle column in each half of the diagram shows the median of the sample in red, where the boxplot indicates the 0.25 and 0.75 percentiles with whiskers extending out no further than 1.5 times the inter-quartile range. The right column indicates the weighted average and standard deviation (in parentheses) of the sample in blue. Evolutionary masses are calculated using the {\sc Bonnsai} tool (Sect.~\ref{sec:massdependence}). } 
\label{fig:massplot}
\end{figure}

Figure~\ref{fig:massplot} presents a summary of the mass properties of the low gravity stars in Fig.~\ref{fig:hunterplotgroups}, comparing the stars in Box\,2 to the rest of the
sample. Two types of statistics are provided, one based on median statistics (Tuley's five-number
summary) and one based on the mean of the sub-samples. Each of these shows
that the objects in Box\,2 are appreciably more massive than the remainder of the stars:
the median evolutionary mass of stars in Box\,2 is 34.8\,\msun, while that of the remaining
stars is 25.8\,\msun.

\citet{brott2011b} performed a similar analysis as done here for a sample of 107 
main-sequence B-type stars in the LMC with projected spin rates up to $\sim$300 km\,s$^{-1}$ 
discussed by \citet{hunter2008,hunter2009}. They too note that the nitrogen enhanced slowly
spinning stars appear to have higher masses. Though the bulk of their sample has a mass
$\leq 12$\,\msun\, the stars that populate Box\,2 almost all have masses in between 
12$-$20\,\msun. Further on in this section we discuss scenarios regarding the nature
of the Box\,2 stars that may explain this behaviour.

\subsection{Current surface nitrogen abundances as predicted by single star evolutionary tracks}
\label{sec:bonnsaicomp}
In addition to the posterior probability distribution of the current mass, {\sc Bonnsai}
also provides the distribution for the current surface nitrogen abundance. For all but one of the stars for
which our analysis resulted in upper limits on \nabun\ only, the upper limit is in 
agreement with the \nabun\ predicted by {\sc Bonnsai}. 
Out of the 38 stars for which we have a direct \nabun\ measurement, only 7 agree within 1$\sigma$ with the nitrogen abundance expected from their evolutionary state. Interestingly, among these are three binaries (namely VFTS\,333, 399 and 440). In the other 31 cases our nitrogen measurement is higher than that predicted by {\sc Bonnsai}. Since all direct measurements have low \vrot, this is an alternative way of expressing that the bulk of our nitrogen rich, slowly rotating sources are not expected from the evolutionary sequences by \citet{brott2011a}.

We repeated the {\sc Bonnsai} analysis for the 9 sources in Fig.~\ref{fig:hunterplotgroups} that have measured abundances and occupy Box~2. In this run, we added the measured surface abundances of helium
and nitrogen to the observables \Teff, $L$, and \vrot. While in the analysis without these
extra two observables {\sc Bonnsai} returned in all cases a modestly rotating star with most
probable inclination $\sin\,i \gtrsim 0.8$, the new run produced fast rotators seen almost
pole-on ($\sin i \lesssim 0.5$) in 8 out of 9 cases. This reflects that in order to reproduce
the high He- and N-abundances in the tracks of \citet{brott2011a} more mixing is required, demanding more rapid rotation. Assuming inclinations randomly distributed 
in 3D space, the probability that out of the 27 stars with $\logg_c \leq 3.75$, 8 have 
$\sin i \leq 0.5$ is 0.015.\footnote{Assuming randomly distributed spin axes in space, 
the probability for an individual system having $\sin\,i \leq x$ is given by $P(x)=1-\sqrt{1-x^2}$. 
As such, the probability for a population of $n$ stars containing $k$ stars with $\sin\,i\leq x$ 
is given by $\frac{n!}{k! (n-k)!} P(x)^k \left(1-P(x)\right)^{n-k}$.} This too demonstrates that
the stars in Box\,2 cannot be straightforwardly understood in the context of current single-star evolutionary 
tracks.

\subsection{Correlation between nitrogen and helium abundance}
\label{sec:NvsHe}

\begin{figure}[t!]
\resizebox{1.05\hsize}{!}{\includegraphics{\figpath 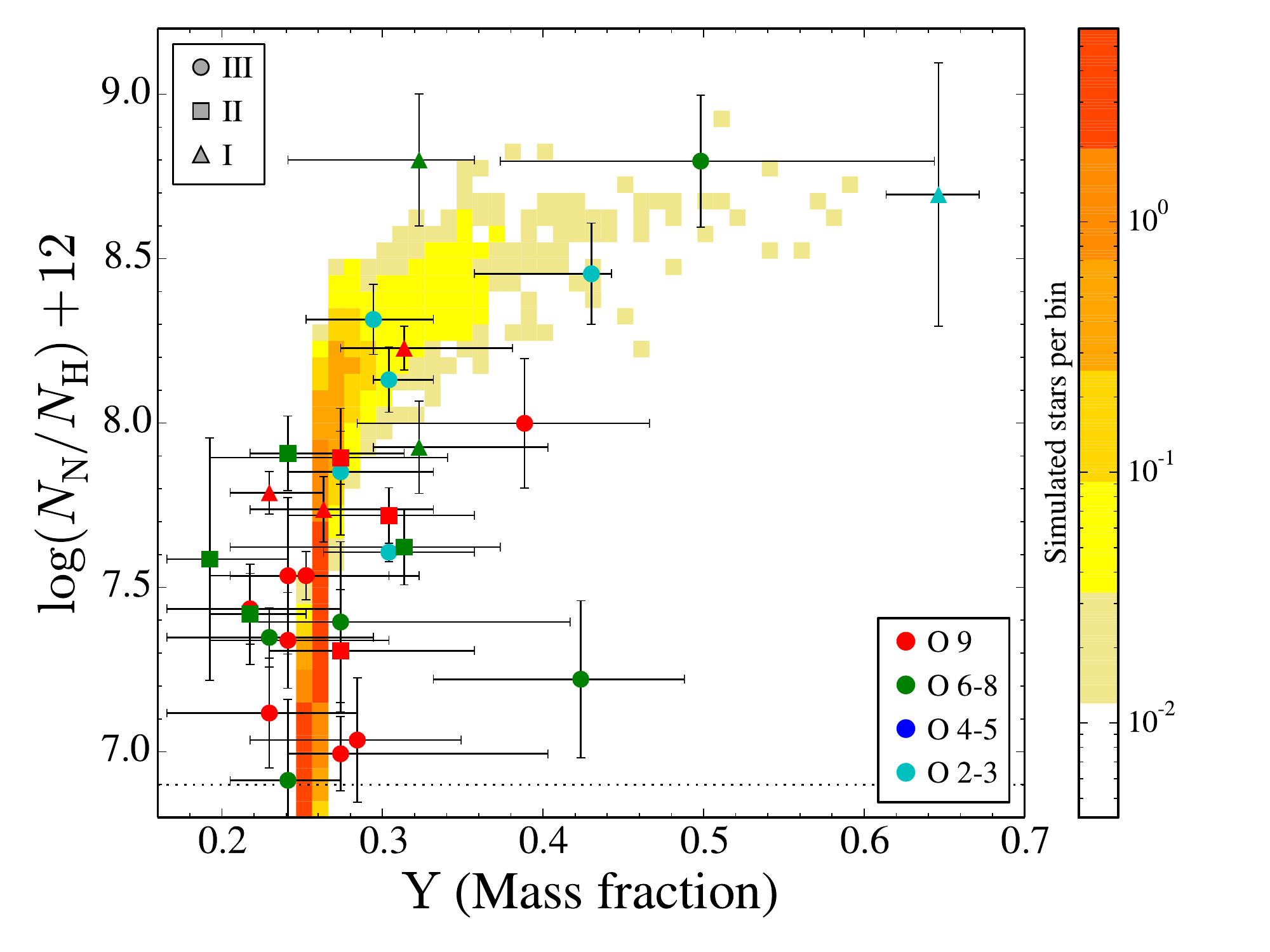}}
\caption{Nitrogen abundance versus surface helium mass fraction. In the background the outcome of 
population synthesis is projected, as explained in Sect.~\ref{sec:popsyn} and shown in 
Fig.~\ref{fig:hunterplotgroups}} 
\label{fig:hunterplotNvsHe}
\end{figure}

In Fig.~\ref{fig:hunterplotNvsHe}, we plot the nitrogen surface abundance derived here versus 
the helium surface fraction (by mass) as derived by Ram\'{i}rez-Agudelo et al. (subm.). Shown in the
background is the same population synthesis calculation as discussed in Sect.~\ref{sec:popsyn_lowg}
and shown in Fig.~\ref{fig:hunterplotgroups}, i.e., for O-type sources that have $\log g \leq 3.75$. 
The results are grouped in bins of $0.01\,(Y) \times 0.05\,(\nabun)$ dex. 
Simulated stars that are helium- as well as nitrogen enriched are all rotating initially at $\gtrsim 350\,\kms$. 

Though the scatter is sizable, the N- and He-abundances of the stars in our sample
appear correlated. Such a correlation is also found by \citet{riverogonzalez2012a}. Moreover, 
the trend agrees quite well with the models, even though the number of predicted and observed stars with He, as well as N, enrichment do not agree. This suggests that though the mechanism that
brings N and He to the surface may still be unidentified, it is one that brings a mixture of
gases to (or deposits a mixture of gases on) the surface that has the He/N abundance ratio
that is expected from the CNO process. 

If the exposed material is CNO-processed, an enhancement in N is expected to be accompanied by a depletion of C. An independent verification would thus be to simultaneously measure the surface C abundances of these stars. It should also be noted that several other surface-enrichment mechanisms (e.g., those discussed in Sect.~\ref{sec:alternativescenarios}) are expected to result in the display of CNO-processed material at the surface.

\subsection{Comparison to other \nabun-\vrot\ studies}
\label{sec:otherNstudies}

We limit a comparison to other studies to main sequence OB stars in the LMC, notably to
the work on early-B stars by \citet{hunter2008,hunter2009} and \citet{brott2011b}, and 
on O-type stars by \citet{riverogonzalez2012a}. For an analysis of LMC B-type supergiants
see \citet{hunter2008} and \citet{mcevoy2015}; of Galactic and SMC B-type
stars see \citet{hunter2009}; for that of Galactic O-type stars, see \citet{martins2015,martins2015b,bouret2012}; of SMC O-type dwarfs, see \citet{bouret2013}.

\citet{riverogonzalez2012a} determined the nitrogen abundance of the LMC stars studied by
\citet{mokiem2007} in the context of the VLT-FLAMES Survey of Massive Stars \citep{evans2006}.
This sample of O and early-B stars of all luminosity classes comprises sources in the central 
LH9 and LH10 associations in the young star-forming region N11, augmented by LMC field stars.
Though the authors refrain from a statistical analysis due to the modest sample size, they do compare with theoretical 
predictions. They find an even larger fraction of stars with N-enrichments that appear too large for their (current) rotation rate. In their work, the correlation between He- and N-abundances (their Fig.~8, our Fig.~\ref{fig:hunterplotNvsHe}) also appears present.

It is interesting that the eight O-type objects in their sample that have 
$\log g_c \leq 3.75$ span essentially the same range in nitrogen abundance found here and 
all would appear in Box~2.
The mean \vrot\ (and associated standard deviation) of this sample of eight is $59 \pm 20$\,\kms, which is lower than the 
98 $\pm$ 12\,\kms\ of our sample in Box~2. The main reason for this is that 
\citeauthor{riverogonzalez2012a}\ accounted for macro-turbulent broadening of their lines, which
in the \vrot\ estimates used here was ignored. In the analysis of the full presumed-single
VFTS O star sample, \citet{ramirezagudelo} employed similar techniques to correct for the
effect of macro-turbulent broadening. Using these results, the mean projected spin velocity
of our stars in Box~2 would be 78 $\pm$ 19\,\kms, i.e., within uncertainties consistent with 
the findings of \citet{riverogonzalez2012a}.  

We conclude that the population of N-enriched slow rotators presented in 
\citet{riverogonzalez2012a} is similar in characteristics to the Box~2 population found here. 
Its presence (in the sample studied by \citeauthor{riverogonzalez2012a}) supports our 
statement that the N-enriched slowly spinning sources are not in agreement with predictions 
of rotational mixing in single stars.

\citet{hunter2008}, in their analysis of over 100 B-type main-sequence stars in the VLT-FLAMES Survey of
Massive Stars, identified a population 
of N-enriched intrinsically slow rotators ($\sim$20\% of their 
sample) and a population of relatively unenriched fast rotators (a further $\sim$20\% of
their sample) that both challenge the 
concept of rotational mixing. The stars in the group of slowly rotating stars preferentially have masses in the range of 12\,\msun\ to 20\,\msun\ and reach 
nitrogen abundances \nabun\ $\sim$ 8, so somewhat lower than the $\sim$8.7 that is reached by the
O stars. 

The definition of the box of N-enriched low \vrot\ sources in \citeauthor{hunter2008} is 
somewhat more stringent than in our case, in that all 
have a projected spin velocity $\vrot < 50$\,\kms. \citet{brott2011b} compared the incidence
of these stars to the same models being tested here and found that this region of the
\nabun\ versus \vrot\ diagram contained 15 times as many stars as predicted. Whether the fact
that the nitrogen-enriched B dwarf population spins somewhat slower than the nitrogen-enriched 
slowly rotating O-type stars and reach somewhat lower N-enrichments points to a fundamental 
difference in the nature of this group remains to be investigated. However, given that the 
global similarities between the two groups are substantial a common nature seems plausible. 

For the group of relatively unenriched fast rotators identified by \citet{hunter2008} 
mostly upper limits to the nitrogen abundance could be constrained. By far the bulk of this
population seems compatible with an abundance that is consistent with the LMC baseline value.
The upper limits of our O stars sample are less constraining and it is therefore too
early to tell whether the fast rotators in both samples behave alike.

\subsection{Alternative scenarios}
\label{sec:alternativescenarios}
In this subsection we discuss scenarios that may yield slowly rotating, nitrogen enriched sources (i.e. Box\,2). It may well be possible that several processes are working in parallel, such that sources may occupy Box~2 for independent, or a combination of, reasons.

\subsubsection{Efficiency of rotational mixing}

The efficiency of rotational mixing in the single-star evolutionary models of \citet{brott2011a} is 
calibrated using efficiency factors. Rotationally induced instabilities contribute in full to the turbulent  viscosity \citep{heger2000a}, while their contribution to the total diffusion coefficient (the way in
which all mixing processes are treated in the models of \citeauthor{brott2011a}) is reduced by a factor $f_{\rm c}$. \citet{brott2011a}
calibrated this parameter using the B dwarfs discussed in Sect.~\ref{sec:otherNstudies}. The inhibiting
effect of chemical gradients on the efficiency of rotational mixing processes is regulated by the
efficiency parameter $f_{\mu} = 0.1$, calibrated following \cite{yoon2006}. The steepness of
the correlations shown in Figs.~\ref{fig:hunterplot} and~\ref{fig:hunterplotgroups} is essentially the 
result of the $f_{\rm c}$ calibration for the quoted $f_{\mu}$. This calibration ignored the nitrogen 
enriched slow rotators (the Box~2 sources) and the relatively unenriched rapid rotators. Assuming the
sources in Box~2 reflect a much larger mixing efficiency, e.g. a much increased $f_{\rm c}$,
an explanation would be required as to why two vastly different regimes of mixing efficiencies would occur in massive stars (i.e., one for Box~2 and one for Box~3 stars).

\subsubsection{Envelope stripping through stellar winds}
\label{sec:windstripping}


The finding that the mass of the Box\,2 stars is on average higher
(Sect.~\ref{sec:massdependence}) could mean that these may have endured a more severe
stripping of their envelopes -- revealing chemically enriched layers -- as stars of 
higher mass (or luminosity) are in general expected to have a stronger mass loss. Mass loss can also drain angular momentum from the surface, resulting in spin-down. This scenario
would be in line with the correlation pattern of the nitrogen and helium abundance as
discussed in Sect.~\ref{sec:NvsHe}.

\citet{bestenlehner2015} investigated the stars on the upper main sequence in the VFTS sample.
The authors find a clear correlation between helium mass fraction $Y$ and $\log\,M/\dot{M}$ at
$\log\,M/\dot{M} \gtrsim -6.5$ (their Fig.~14), consistent with the idea that severe mass loss in 
very massive stars indeed reveals the deeper layers. Ram\'{i}rez-Agudelo et al. (subm.)
added the sources studied here to this diagram, populating the regime at $\log\,M/\dot{M}
\lesssim -6.5$. They find that, save for a few outliers, a monotonic trend of $Y$ versus $\log\,M/\dot{M}$ is missing in
this regime\footnote{For completeness we note that one of our sources, VFTS\,180, populates the regime studied
by \citet{bestenlehner2015}. This source does seem to follow the trend reported by
these authors.} (their Fig.~7).

The latter is also expected from the models of \citet{brott2011a}. For example, Fig.~4 of \citeauthor{brott2011a} shows the expected surface enrichment at the TAMS, as a function of initial mass and rotation rate. It can be seen that, at LMC metallicity and typical rotation rates, significant surface enrichment is expected only for stars in excess of $60\,\msun$. Note also that these are values as predicted for the TAMS, while most of our observed stars are still undergoing main-sequence evolution. Furthermore, even though the average mass of Box~2 sources compared to the remaining stars is higher, most stars have evolutionary masses on the order of $\sim30\,\msun$.

The implicit assumption in the above discussion is that the mass-loss rates that
are adopted in the evolutionary tracks being scrutinised here are correct. The $\dot{M}$
recipe used in these models are from \citet{vink2000,vink2001}. These are consistent with the
assumption of moderately clumped winds (\citeauthor{mokiem2007} \citeyear{mokiem2007}; 
Ram\'{i}rez-Agudelo et al. subm.). Hence significantly higher mass-loss rates would 
require the outflows to be close to homogeneous, which is not expected. 

In the context of envelope stripping it would 
be hard to reconcile a mass dependence of Box\,2 versus non-Box\,2 sources in both 
the B stars sample of \citet{hunter2008} and our O star sample, the Box\,2 B-type stars 
having a lower average mass than the non-Box\,2 O-type stars.
Moreover, at typical rotation rates wind stripping is thought to be important only for the most massive stars (as explained above), such that the anomalous B stars of \citeauthor{hunter2008} are unlikely to have resulted from this process. 

We conclude that, although for some stars, envelope stripping could be an important aspect, for the bulk of the Box~2 objects it is unlikely to be the cause of the observed N-enrichment.

\subsubsection{Binary evolution}
\label{sec:binaryevolution}

Binary interaction in general might yield stars of higher mass (as the mass gainer may dominate 
the light of the system, or when two stars yield a more massive single star after a merger event), in accordance with
the observations described in Sect.~\ref{sec:massdependence}.

It is not improbable that our
sample of presumed single stars is polluted with binary products.
\citet{demink2014} showed that using radial velocities variations to select a sample against binaries may inadvertently {\em increase} the fraction of binary interaction products. This is a result of two considerations. First, merger products are essentially single stars. Second, in a typical post mass transfer system the light is dominated by the mass gainer, which in many cases displays only modest radial velocity variations (less than 20\,\kms). Such a population of post-interaction
binaries can explain the presence of a high-velocity tail ($\vrot > 300\,\kms$) found in the
spin distribution of the VFTS single O star sample by \citet{ramirezagudelo}, which was 
proposed by \citet{demink2013}.
Interestingly, this tail appears absent in the spin distribution of relatively wide pre-interacting binaries, i.e., sources that are not affected by tidal effects \citep{ramirezagudelo2015}, supporting the idea that the high-velocity tail of the
spin distribution is the result of binary interaction. So indeed, a sizable fraction of our presumed single-star
sample may in fact be a product of binary evolution.

Several scenarios involving binary interaction might explain the stars in Box~2 (and stars in Box~3 if their nitrogen abundances turn out to be too low to concord with rotational mixing). 
\citet{glebbeek2013}, for instance, investigate the evolution of merger remnants. They find that mixing of the envelopes, for example through thermohaline circulation, can result in enhanced nitrogen abundances. These enhancements are stronger for more massive stars.
The possibility of Box\,2 stars being binary products has also been brought forward by \citet{brott2011b} to explain the B-dwarfs
analysed by \citet{hunter2008}, to which we refer for a more extended discussion. 
%
%
Note, however, that binary interaction through mass transfer is more likely to produce fast rotators, be they enriched or not. Channels reproducing the slow rotators in Box~2, for example through spin-down by resonant locking, are relatively less likely with respect to channels reproducing rapid rotators.


A quantitative test of the binary hypothesis requires binary population synthesis models of rotationally mixed stars, which are not yet available
\citep[but see][for a discussion of first steps in this direction]{langer2012}. An empirical approach to probing a post-interaction nature of the stars in Box~2 is to compare the nitrogen abundances of the sample studied here to those of the binaries in the VFTS.

It should be noted that if binarity is at play, one must consider the contribution of a possibly
undetected companion to the total observed continuum light. This contribution may dilute the
strength of the nitrogen lines, implying the derived N-abundances reported here would formally
represent lower limits in those cases.

\subsubsection{Magnetic fields}

Magnetic fields have been suggested to play a role in explaining the N-enhanced,
slowly rotating early-B dwarfs that challenge rotational mixing \citep{morel2008,morel2012,przybilla2011}. 
Such a field  may be either of fossil origin or due to a rotationally driven dynamo 
operating in the radiative zone of the star \citep{spruit1999,maeder2004}.

\citet{meynet2011} proposed that magnetic braking \citep[see][]{udDoula2002,udDoula2008,udDoula2009} 
during the main-sequence phase leads to slow rotation. In their model, \citeauthor{meynet2011}
assume the absence of a magnetic field in the stellar interior. Consequently, they find that
the slowing down of the surface layers results in a strong differential rotation, resulting in
a nitrogen surface enhancement. A situation where the observable magnetosphere
is the external part of a strong toroidal-poloidal field inside the star
\citep{braithwaite2004} might induce considerably less mixing \citep[see also][]{morel2012}.

\citet{potter2012} discussed the $\alpha-\Omega$ dynamo as a mechanism for driving the
generation of large-scale magnetic flux and found a strong mass dependence for the dynamo-driven
field. For stars with initial masses greater than about 15 \msun\ (i.e., those born as O stars),
they find that the dynamo cannot be sustained. Initially lower mass stars (i.e., those born as B or later
type stars) with sufficiently high rotation rates are found to develop an active dynamo and so exhibit
strong magnetic fields. They are spun down quickly by magnetic braking and
magneto-rotational turbulence \citep{spruit2002} causing changes to the surface
composition. Our finding that also the O stars \citep[in addition to the B stars; see][]{hunter2008}
populate Box\,2 might thus indicate that if magnetic fields play a role in explaining the nature of
these N-enriched slowly spinning stars their fields are of fossil origin. If not, two
different dynamo models may need to be invoked to explain the Box\,2 objects, one
for O-type stars and another for B-type stars.

Alternatively, magnetic fields may play a role in conjunction with binary evolution. If the mass gainer is  spun up by angular momentum transfer, efficient rotational mixing leads to the surfacing of nitrogen. It may generate magnetic fields that, after the surface is enhanced in nitrogen, spin down the star as a result of magnetic braking.
As an extreme example of this mechanism one may envision the actual merger of the two
stars. Such merger products may represent of the order of 10\% of the O star field
population \citep{demink2014}, a similar percentage as the incidence rate of magnetic O-type stars \citep{grunhut2012,wade2014,fossati2015}. This is, however, less than the percentage of stars found in Box~2.

\section{Summary}
\label{sec:conclusions}

We have determined the nitrogen abundances of the O-type giants and supergiants observed in the
context of the VLT-FLAMES Tarantula Survey (VFTS) and compared these to evolutionary models of rotating
stars with the aim to quantitatively test rotational mixing at the hot and bright end of the
main-sequence. Using stellar parameters determined by Ram\'{i}rez-Agudelo et al. (subm.), we 
estimate the surface nitrogen abundances of stars by comparing synthetic equivalent widths
of a set of (weak) optical N\,{\sc ii-v} lines to those measured from the spectra. Of the 72 sources 
in our sample, we can constrain the nitrogen abundance for 38. For 34 sources we cannot 
distinguish the signature of the shallow lines from the noise and we have to settle for upper limits
only. All but one of the stars with projected rotational velocities above $170\,\kms$ are in this latter group.


Rotationally induced mixing transports CNO-processed material from the stellar interior to the 
surface. Predictions show that for larger initial rotational velocity, the speedier the N-enrichment sets in and the 
higher the N-abundance that is reached at the end of the main-sequence \citep{brott2011a,ekstrom2012}.
We performed a quantitative test of the predictions for O-type stars with (rotation corrected) surface gravities 
$\logg_c \leq 3.75$, for which our sample is near to complete within the VFTS.
From a comparison of the behaviour of \nabun\ as a function of \vrot\ with population
synthesis computations, assuming a continuous formation of single stars we conclude the
following:

\begin{itemize}
\item The nitrogen surface abundance of 60$-$70 percent of the $\logg_c \leq 3.75$ stars in our sample is compatible with the predictions of rotational mixing as implemented by \citet{brott2011a}. The upper limits that we derive for stars with $\vrot \gtrsim 170\,\kms$ reveal no apparent contradiction; obtaining higher signal-to-noise ratio data of these fast rotating stars is however desirable to further confront the predictions of rotational mixing theory at large \vrot.

\item About 30$-$40 percent of the low-gravity sample displays projected spin rates less than
      140\,\kms\ whilst showing strong N-enhancement. This group (the Box~2 stars in 
      Fig.~\ref{fig:hunterplotgroups}) contains almost 5 times as many sources as expected and is 
      not compatible with predictions of rotational 
      mixing. A similar group of N-enhanced slowly spinning objects defying rotational mixing
      theory has been identified in an LMC population of B-type dwarf stars \citep{hunter2008}.
    
\item The mean projected spin velocity of the N-enhanced slowly spinning sources is
      $78 \pm 19$\,\kms\ when correcting for line broadening by macro-turbulent motions.
      The strongly enriched slow rotators studied by \citet{hunter2008} all have 
      $\vrot < 50$\,\kms. Whether this difference in projected spin rate points to a
      different nature for these two groups remains to be investigated.
      
\item The mean evolutionary mass of the N-enhanced slowly spinning sources is $38 \pm 13$\,\msun,
      which is somewhat larger than the $28 \pm 8$\,\msun\ of the remaining stars.
      
\item The correlation between the N and He abundance of the full set of O-type giants to
      supergiants studied here is compatible with expectations of rotational mixing. This
      implies that the mechanism that enriches the surface -- be it rotational mixing
      or something else -- transports material that is a gas mixture that is expected
      to arise during the internal evolution of stars. This hypothesis could be verified by future measurements of the C abundance of this sample.
\end{itemize}


The present study has focused on presumably single giants and supergiants O-type stars in the VFTS. The nitrogen surface abundance of about 2/3 of the data are compatible with the expectation of rotational mixing, while the remaining third show a N-abundance that is too large for their projected spin rate. It remains to be investigated whether effects of binarity and/or magnetic fields need to be invoked to explain this anomalous group of stars. One way forward is to perform population synthesis that accounts for the effect of binarity and, possibly, magnetic fields. Another way is to measure the nitrogen abundances of (O-type) binaries. If the anomalous nitrogen surface abundance of slow rotators is the result of binary interaction through mass transfer and/or coalescence, one may expect such a group to be largely absent in (pre-interaction) spectroscopic binaries.



\begin{acknowledgements}
N.J.G. is part of the International Max Planck Research School (IMPRS) for Astronomy and Astrophysics at the Universities of Bonn and Cologne.
\end{acknowledgements}

\bibliographystyle{aa}	

\begin{appendix}

\section{Full table of results}

\subsection{O-type giants and supergiants}
\label{app:datatable}

\onecolumn

%
\begin{landscape}
\begin{center} 

\setlength\LTcapwidth{\textwidth}
\begin{longtable}[H]{l l r c c c c c c c c}

\caption{Nitrogen abundances of presumed-single O-type giants and supergiants in the VFTS. Upper limits are indicated by $<$ in the fifth column and the number of lines used for the measurement $n_l$ is given in the sixth column. Abundances have been estimated with fixed $\vmicro=10\,\kms$ (see Sect.~\ref{sec:limitations}). Key parameters that are used in this paper ($\vrot$, $\mathrm{Y}$, $\logg_c$, $\Teff$, $\logL$, $\mathrm{M}_\mathrm{evol}$) were determined by Ram\'{i}rez-Agudelo et al. (subm.). Newly detected binaries are shown in a separate segment; see Sect.~\ref{sec:limitations} for a discussion.}
\label{tab:nresults} \\ 
\hline
\hline\\[-9pt]
VFTS & SpT$^a$ & $\vrot$ & \nabun &  $\sigma_{\mathrm{N}}^b$ & $n_\mathrm{l}$ & $\mathrm{Y}^c$ & $\logg_c^d$& $\Teff$ & $\logL$ & $\mathrm{M}_\mathrm{evol}^e$ \\[2pt] 
 & & \kms & dex & dex & & & cgs & $\mathrm{kK}$ & $L_\odot$ & $\mathrm{M}_\odot$ \\[2pt] 
\hline\\[-9pt] 
\endfirsthead
\hline
\hline\\[-9pt]
VFTS & SpT$^a$ & $\vrot$ & \nabun &  $\sigma_{\mathrm{N}}^b$ & $n_\mathrm{l}$ & $\mathrm{Y}^c$ & $\logg_c^d$& $\Teff$ & $\logL$ & $\mathrm{M}_\mathrm{evol}^e$ \\[2pt] 
 & & \kms & dex & dex & & & cgs & $\mathrm{kK}$ & $L_\odot$ & $\mathrm{M}_\odot$ \\[2pt] 
\hline\\[-9pt] 
\endhead
\hline
\endfoot

\hline\\[-9pt]
\multicolumn{11}{p{0.8\textwidth}}{{\bf Notes.} (a) Spectral types as given by \citet{walborn2014}; (b) The errors are computed from the standard deviation between measurements from individual lines. They do not incorporate the uncertainties in the atmospheric parameters and continuum normalisation (see Sect.~\ref{sec:method}). (c) $\mathrm{Y}$ is the surface helium mass fraction; (d) $\logg_c$ (cgs) is the logarithmic surface gravity corrected for centrifugal force; (e) evolutionary masses $M_\mathrm{evol}$ are determined with {\sc Bonnsai} (Sect.~\ref{sec:massdependence}).}
\endlastfoot
016 & O2 III-If* & 112 & 7.85 & 0.19 & 3 & 0.27 & 4.03 & 50.60 & 6.12 & 94 \\
035 & O9.5 IIIn & 346 & 8.62 & $<$ & $-$ & 0.24 & 4.27 & 32.55 & 4.37 & 16 \\
046 & O9.7 II((n)) & 168 & 7.57 & $<$ & $-$ & 0.39 & 3.30 & 28.85 & 5.09 & 21 \\
070 & O9.7 II & 126 & 8.08 & $<$ & $-$ & 0.25 & 4.23 & 32.15 & 4.47 & 15 \\
076 & O9.2 III & 90 & 6.99 & 0.11 & 1 & 0.27 & 3.56 & 33.25 & 5.10 & 23 \\
077 & O9.5 :IIIn & 264 & 8.58 & $<$ & $-$ & 0.17 & 4.32 & 33.65 & 4.48 & 17 \\
080 & O9.7 II-III((n)) & 194 & 8.33 & $<$ & $-$ & 0.28 & 3.89 & 31.30 & 4.68 & 17 \\
087 & O9.7 Ib-II & 84 & 7.79 & 0.06 & 6 & 0.23 & 3.32 & 30.55 & 5.29 & 27 \\
091 & O9.5 IIIn & 308 & 8.05 & $<$ & $-$ & 0.27 & 3.98 & 32.50 & 4.79 & 19 \\
103 & O8.5 III((f)) & 126 & 7.35 & 0.09 & 3 & 0.23 & 3.89 & 34.70 & 5.21 & 26 \\
104 & O9.7 II-III((n)) & 198 & 7.99 & $<$ & $-$ & 0.23 & 4.07 & 30.80 & 4.31 & 14 \\
109 & O9.7 II:n & 352 & 7.99 & $<$ & $-$ & 0.29 & 3.66 & 24.35 & 4.25 & 11 \\
113 & O9.7 II or B0 IV? & 12 & 7.75 & $<$ & $-$ & 0.24 & 4.47 & 33.30 & 4.46 & 16 \\
128 & O9.5 III:((n)) & 180 & 7.89 & $<$ & $-$ & 0.28 & 4.26 & 33.80 & 4.46 & 17 \\
141 & O9.5 II-III((n)) & 166 & 7.96 & $<$ & $-$ & 0.26 & 4.26 & 32.00 & 4.82 & 18 \\
151 & O6.5 II(f)p & 118 & 7.91 & 0.11 & 3 & 0.24 & 3.95 & 37.65 & 5.87 & 53 \\
153 & O9 III((n)) & 158 & 7.53 & $<$ & $-$ & 0.24 & 4.17 & 35.50 & 5.22 & 27 \\
160 & O9.5 III((n)) & 162 & 7.18 & $<$ & $-$ & 0.32 & 3.66 & 32.30 & 5.36 & 29 \\
172 & O9 III((f)) & 118 & 7.34 & 0.15 & 4 & 0.24 & 3.88 & 34.70 & 4.50 & 18 \\
178 & O9.7 Iab & 90 & 7.74 & 0.10 & 7 & 0.26 & 3.18 & 28.25 & 5.60 & 36 \\
180 & O3 If* & 118 & 8.70 & 0.40 & 6 & 0.65 & 3.44 & 40.45 & 5.85 & 53 \\
185 & O7.5 III((f)) & 136 & 6.91 & 0.25 & 2 & 0.24 & 3.40 & 34.50 & 5.28 & 26 \\
188 & O9.7 :III: & 126 & 8.33 & $<$ & $-$ & 0.18 & 4.51 & 33.65 & 4.66 & 18 \\
192 & O9.7 II or B0 IV? & 46 & 7.14 & $<$ & $-$ & 0.24 & 4.19 & 31.30 & 4.30 & 14 \\
205 & O9.7 II((n)) or B0 IV((n))? & 158 & 7.67 & $<$ & $-$ & 0.26 & 4.32 & 30.20 & 4.46 & 14 \\
207 & O9.7 II((n)) & 166 & 7.84 & $<$ & $-$ & 0.25 & 4.31 & 30.80 & 4.42 & 15 \\
210 & O9.7 II-III((n)) & 162 & 7.64 & $<$ & $-$ & 0.29 & 4.07 & 32.30 & 4.60 & 17 \\
226 & O9.7 III & 64 & 7.54 & 0.24 & 2 & 0.24 & 4.25 & 32.30 & 4.43 & 16 \\
235 & O9.7 III & 18 & 7.38 & $<$ & $-$ & 0.30 & 4.08 & 32.30 & 4.62 & 17 \\
244 & O5 III(n)(fc) & 230 & 8.26 & $<$ & $-$ & 0.29 & 3.71 & 41.05 & 5.58 & 39 \\
253 & O9.5 II & 96 & 7.72 & 0.08 & 3 & 0.30 & 4.09 & 30.95 & 4.85 & 18 \\
259 & O6 Iaf & 92 & 8.80 & 0.20 & 4 & 0.32 & 3.49 & 36.80 & 6.00 & 63 \\
267 & O3 III-I(n)f* & 182 & 7.61 & 0.03 & 2 & 0.30 & 3.90 & 44.10 & 5.96 & 66 \\
304 & O9.7 III & 10 & 7.15 & $<$ & $-$ & 0.21 & 4.18 & 31.60 & 4.34 & 15 \\
306 & O8.5 II((f)) & 90 & 7.62 & 0.12 & 5 & 0.31 & 3.27 & 31.50 & 5.36 & 29 \\
328 & O9.5 III(n) & 244 & 8.22 & $<$ & $-$ & 0.29 & 4.23 & 33.25 & 4.45 & 17 \\
346 & O9.7 III & 92 & 7.43 & 0.11 & 2 & 0.22 & 4.23 & 31.70 & 4.56 & 16 \\
370 & O9.7 III & 84 & 7.64 & $<$ & $-$ & 0.24 & 4.14 & 32.65 & 4.54 & 16 \\
466 & O9 III & 88 & 7.54 & 0.07 & 4 & 0.25 & 3.59 & 33.80 & 5.22 & 26 \\
495 & O9.7 II-IIIn & 218 & 8.22 & $<$ & $-$ & 0.25 & 4.33 & 31.45 & 4.55 & 16 \\
502 & O9.7 II & 102 & 7.31 & 0.19 & 5 & 0.27 & 3.27 & 29.75 & 5.55 & 32 \\
503 & O9 III & 90 & 7.04 & 0.19 & 4 & 0.28 & 3.40 & 32.10 & 5.08 & 22 \\
513 & O6-7 II(f) & 130 & 7.59 & 0.37 & 2 & 0.19 & 4.21 & 39.05 & 5.00 & 26 \\
518 & O3.5 III(f*) & 112 & 8.45 & 0.15 & 5 & 0.43 & 3.67 & 44.85 & 5.67 & 48 \\
546 & O8-9 III:((n)) & 94 & 7.22 & 0.24 & 2 & 0.42 & 3.46 & 31.60 & 4.94 & 19 \\
566 & O3 III(f*) & 128 & 8.32 & 0.11 & 8 & 0.29 & 3.77 & 45.70 & 5.83 & 61 \\
569 & O9.2 III: & 48 & 7.12 & 0.17 & 1 & 0.23 & 3.87 & 32.55 & 4.74 & 18 \\
571 & O9.5 II-III(n) & 148 & 7.70 & $<$ & $-$ & 0.24 & 4.31 & 31.10 & 4.39 & 15 \\
574 & O9.5 IIIn & 270 & 8.29 & $<$ & $-$ & 0.28 & 4.11 & 31.40 & 4.36 & 15 \\
599 & O3 III(f*) & 130 & 8.13 & 0.10 & 3 & 0.30 & 4.02 & 47.30 & 6.01 & 69 \\
607 & O9.7 III & 60 & 7.64 & $<$ & $-$ & 0.19 & 4.23 & 32.80 & 4.56 & 16 \\
615 & O9.5 IIInn & 372 & 8.52 & $<$ & $-$ & 0.18 & 4.08 & 30.70 & 4.92 & 19 \\
620 & O9.7 III(n) & 208 & 8.35 & $<$ & $-$ & 0.28 & 4.11 & 31.70 & 4.31 & 15 \\
622 & O9.7 III & 90 & 7.83 & $<$ & $-$ & 0.24 & 4.31 & 31.20 & 4.25 & 14 \\
664 & O7 II(f) & 98 & 7.42 & 0.15 & 3 & 0.22 & 3.58 & 35.70 & 5.53 & 36 \\
669 & O8 Ib(f) & 112 & 7.93 & 0.14 & 5 & 0.32 & 3.25 & 33.30 & 5.51 & 35 \\
711 & O9.7 III & 39 & 7.62 & $<$ & $-$ & 0.18 & 4.47 & 32.80 & 4.73 & 18 \\
753 & O9.7 II-III & 30 & 7.89 & 0.08 & 3 & 0.27 & 4.14 & 33.30 & 4.81 & 19 \\
764 & O9.7 Ia & 92 & 8.23 & 0.07 & 7 & 0.31 & 2.90 & 28.85 & 5.39 & 28 \\
777 & O9.2 II & 138 & 7.71 & $<$ & $-$ & 0.24 & 3.19 & 29.30 & 5.30 & 26 \\
782 & O8.5 III & 82 & 7.39 & 0.24 & 3 & 0.27 & 3.47 & 33.80 & 5.20 & 26 \\
787 & O9.7 III & 56 & 7.97 & $<$ & $-$ & 0.23 & 4.45 & 33.25 & 4.55 & 17 \\
807 & O9.5 III & 28 & 8.00 & 0.20 & 5 & 0.39 & 3.77 & 33.25 & 4.83 & 19 \\
819 & ON8 III((f)) & 70 & 8.80 & 0.20 & 4 & 0.50 & 3.82 & 36.65 & 4.86 & 22 \\
843 & O9.5 IIIn & 318 & 8.26 & $<$ & $-$ & 0.28 & 4.02 & 30.50 & 4.44 & 16 \\
\hline\\[-9pt] 
\multicolumn{11}{c}{Newly detected spectroscopic binaries}\\
\hline\\[-9pt] 
064 & O7.5 II(f) & 116 & 7.93 & 0.12 & 4 & 0.21 & 3.72 & 35.65 & 5.74 & 44 \\
093 & O9.2 III-IV & 64 & 7.60 & 0.10 & 5 & 0.28 & 3.88 & 34.50 & 5.02 & 22 \\
171 & O8 II-III(f) & 92 & 7.48 & 0.14 & 5 & 0.27 & 3.56 & 34.25 & 5.43 & 32 \\
332 & O9.2 II-III & 84 & 7.32 & 0.07 & 5 & 0.28 & 3.46 & 32.25 & 5.32 & 27 \\
333 & O8 II-III((f)) & 114 & 7.48 & 0.14 & 3 & 0.24 & 3.46 & 33.80 & 5.88 & 50 \\
399 & O9 IIIn & 324 & 7.67 & 0.10 & 1 & 0.22 & 3.54 & 30.10 & 4.81 & 17 \\
440 & O6-6.5 II(f) & 146 & 7.63 & 0.13 & 5 & 0.26 & 3.31 & 33.80 & 5.63 & 39 \\

\end{longtable}
\end{center}
\end{landscape}

\twocolumn

\subsection{O-type stars with no luminosity class}
\label{app:datatablenoLC}

\begin{figure}[!th]
\centering
\resizebox{\hsize}{!}{\includegraphics{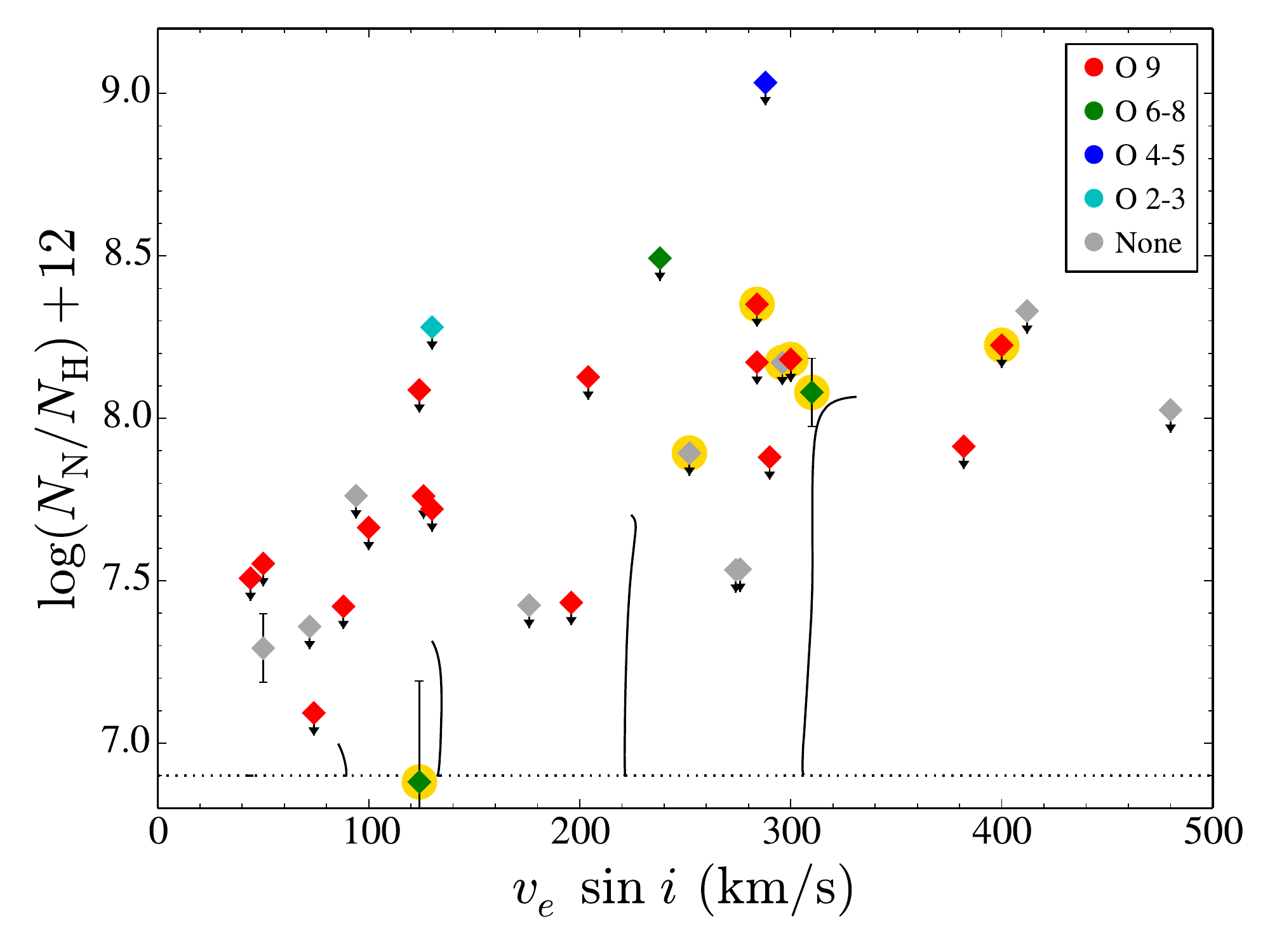}}
\caption{Nitrogen abundance versus projected rotational velocity for the stars in the VFTS sample with no luminosity class identifier. Colours denote spectral type, where grey indicates no identified spectral type \citep{walborn2014}. Stars with a yellow circle underlaid are considered poor quality fits.}
\label{fig:hunterplotnoLC}
\end{figure}

Here we present constraints on the nitrogen abundances of the 31 VFTS O-type stars for which no luminosity class identifier has been assigned by \citet{walborn2014}. The nitrogen constraints are listed in Table~\ref{tab:nresultsnoLC} and visualised in Fig.~\ref{fig:hunterplotnoLC}. 
For 6 sources, Ram\'{i}rez-Agudelo et al. (subm.) find no satisfactory fit to the spectral lines. For one source, VFTS\,177, nitrogen lines appear present but spurious emission at the location of \ion{N}{iii}\,$\lambda4511$ prevents a reliable extraction of the nitrogen abundance of the combined (blended) \ion{N}{iii}\,$\lambda$4511-4515-4518 complex. The 7 aforementioned sources have a yellow circle underlaid in Fig.~\ref{fig:hunterplotnoLC} and are excluded from the discussion in Sect.~\ref{sec:popsyn_lowg}.\\[0.4cm]



%
\tablehead{
\hline\hline\\[-9pt]
VFTS & SpT$^a$ & $\vrot$ & \nabun &  $\sigma_{\mathrm{N}}$ & $n_\mathrm{l}$ & $\logg_c^b$ \\[2pt] 
 & & \kms & dex & dex & & cgs \\[2pt] 
\hline\\[-9pt] 
}
\topcaption{Nitrogen abundances of presumed-single O-type sources with no luminosity class identifier. Upper limits are indicated by $<$ in the fifth column and the number of lines used for the measurement $n_l$ is given in the sixth column. Abundances have been estimated with fixed $\vmicro=10\,\kms$ (see Sect.~\ref{sec:limitations}). Poor quality fits are shown in a separate segment.} 

\tabletail{\hline}
\tablelasttail{\hline\\[-9pt]
\multicolumn{7}{p{0.95\linewidth}}{{\bf Notes.} (a) Spectral types as given by \citet{walborn2014}; (b) $\logg_c$ (cgs) is the logarithmic surface gravity corrected for centrifugal force.}
}

\begin{supertabular}{l l r c c c c}\label{tab:nresultsnoLC} 
051 & OBpe & 412 & 8.33 & $<$ & $-$ & 3.44 \\
125 & Ope & 274 & 7.53 & $<$ & $-$ & 4.04 \\
131 & O9.7 & 124 & 8.09 & $<$ & $-$ & 4.59 \\
142 & Op & 72 & 7.36 & $<$ & $-$ & 4.22 \\
208 & O6(n)fp & 238 & 8.49 & $<$ & $-$ & 3.61 \\
373 & O9.5n & 382 & 7.91 & $<$ & $-$ & 3.83 \\
393 & O9.5(n) & 196 & 7.43 & $<$ & $-$ & 3.55 \\
405 & O9.5:n & 290 & 7.88 & $<$ & $-$ & 3.85 \\
412 & O9.7 & 50 & 7.55 & $<$ & $-$ & 4.08 \\
444 & O9.7 & 100 & 7.66 & $<$ & $-$ & 4.23 \\
456 & Onn & 480 & 8.03 & $<$ & $-$ & 3.93 \\
465 & On & 276 & 7.54 & $<$ & $-$ & 3.77 \\
476 & O((n)) & 176 & 7.42 & $<$ & $-$ & 3.31 \\
477 & O((n)) & 94 & 7.76 & $<$ & $-$ & 3.88 \\
515 & O8-9p & 50 & 7.29 & 0.11 & 2 & 3.89 \\
519 & O3-4((f)) & 130 & 8.28 & $<$ & $-$ & 3.66 \\
528 & O9.7(n) & 130 & 7.72 & $<$ & $-$ & 4.14 \\
529 & O9.5(n)SB? & 284 & 8.17 & $<$ & $-$ & 4.34 \\
539 & O9.5(n) & 126 & 7.76 & $<$ & $-$ & 4.08 \\
559 & O9.7(n) & 204 & 8.13 & $<$ & $-$ & 4.20 \\
579 & O9:((n))SB? & 88 & 7.42 & $<$ & $-$ & 3.94 \\
587 & O9.7:SB? & 74 & 7.09 & $<$ & $-$ & 4.31 \\
594 & O9.7 & 44 & 7.51 & $<$ & $-$ & 4.20 \\
626 & O5-6n(f)p & 288 & 9.03 & $<$ & $-$ & 3.70 \\
\hline\\[-9pt] 
\multicolumn{7}{c}{Poor quality fits}\\
\hline\\[-9pt] 
145 & O8fp & 124 & 6.88 & 0.31 & 2 & 3.57 \\
177 & O7n(f)p & 310 & 8.08 & 0.10 & 1 & 3.66 \\
360 & O9.7 & 400 & 8.23 & $<$ & $-$ & 3.37 \\
400 & O9.7 & 284 & 8.35 & $<$ & $-$ & 4.30 \\
446 & Onn((f)) & 252 & 7.89 & $<$ & $-$ & 3.48 \\
451 & O(n) & 296 & 8.17 & $<$ & $-$ & 3.79 \\
565 & O9.5:SB? & 300 & 8.18 & $<$ & $-$ & 4.32 \\
\end{supertabular}

\newpage
\section{Discussion on individual sources}
\label{app:individuals}

In this paper, we discuss the sample of VFTS O-type giants and supergiants from the viewpoint of a coherent, continuously evolving stellar population (Sect.~\ref{sec:popsyn}). However, given the spatial and temporal extent of the 30\,Dor region covered by the VFTS data \citep[see e.g.,][for a discussion]{walborn1997,selman1999,sabbi2013}, 
one has to be aware of the diversity within the sample. 

We provide here more details on individual stars, that have $\nabun>7.5$ or are otherwise noteworthy, such as the run-away candidates. We discuss their evolutionary state and notable aspects of the spectral fitting,  supplemented by findings of \citet{walborn2014} and Almeida et al. (in prep.). \citeauthor{walborn2014} presented the spectral classification of this sample, and Almeida et al. present the results of the follow-up radial velocity study on the VFTS O-type stars. This section is organised as follows. We first discuss some general properties of the sample. Next, we discuss the run-away candidates, followed by the confirmed binaries, stars with radial velocity variations that do not resemble a binary orbit, confirmed single stars, and end with stars not included in the sample of Almeida et al. (in prep.).

\begin{figure}[t]
\centering
\resizebox{\hsize}{!}{\includegraphics{\figpath 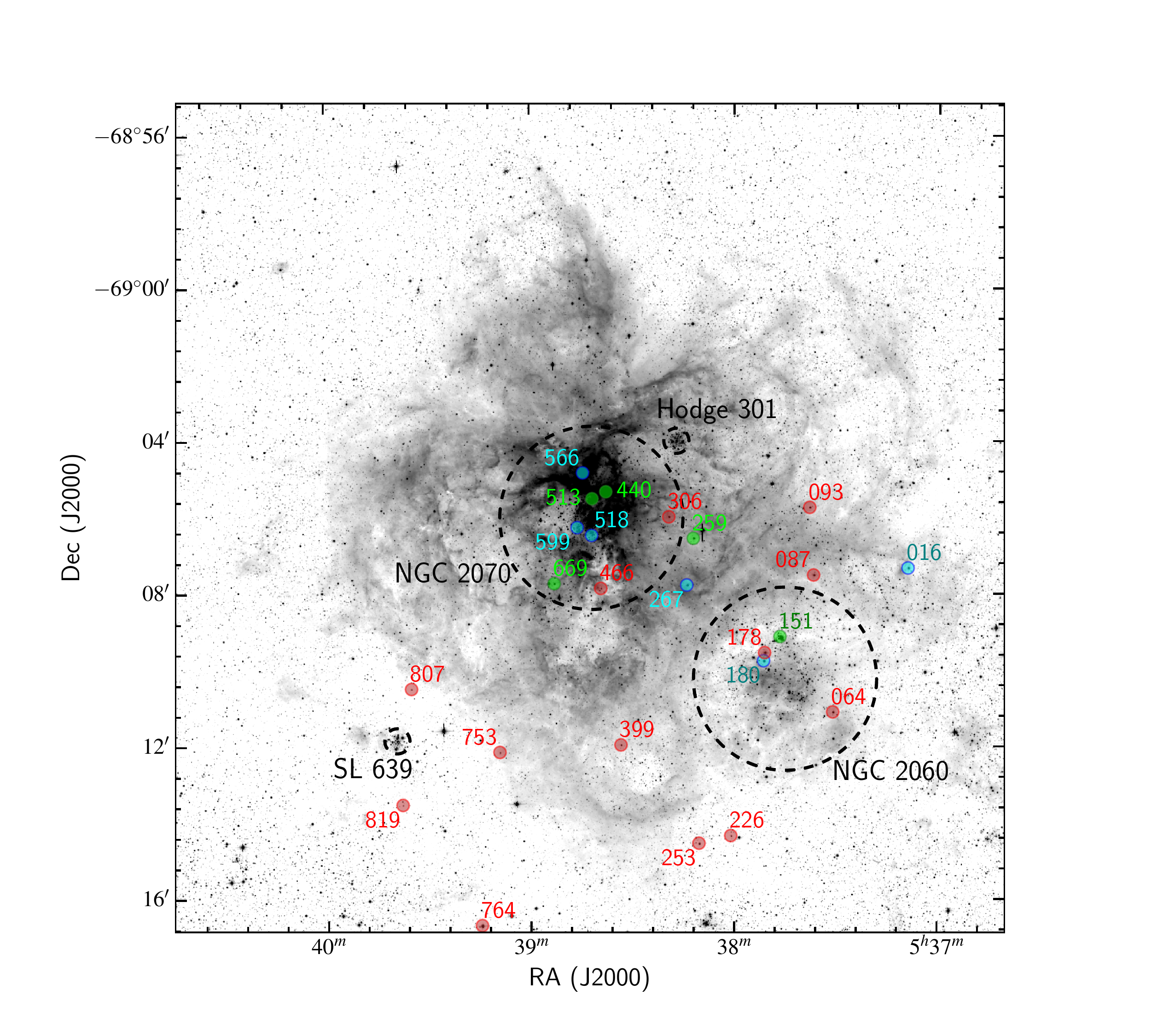}}
\caption{Spatial distribution of the O-type giants and supergiants with measured $\nabun>7.5$. The locations of the different clusters are indicated. Spectral types O\,2-3 are indicated in blue, the O\,6-7 in green, and sources  of spectral type O\,8 and later are coloured red.}
\label{fig:spatial}
\end{figure}

{\it General properties:} In Fig.~\ref{fig:spatial} we display the location of stars with measured $\nabun>7.5$ (excluding upper limits, essentially Box~2 stars). Members of this group are found in the field, as well as in the different clusters.

The luminous stars in our sample are on average N-enriched, as can be seen in Fig.~\ref{fig:HRD}. Moreover, the stars that show He- as well as N-enrichment (Fig.~\ref{fig:hunterplotNvsHe}) are on average luminous. Furthermore, it appears that the O2-3 objects are all enhanced in nitrogen to a certain degree. 

{\it Run-away candidates:} VFTS~016, 226 and 306 are classified as run-away candidates. The three stars are moderately rotating with \vrot\ on the order of $100\,\kms$ and, interestingly, all occupy the Box~2 region (Fig.~\ref{fig:hunterplotgroups}). The hot star VFTS~016 was also studied by \citet{evans2010b}, who estimated a nitrogen abundance of $\nabun \sim 7.9$, in good agreement with our result. Its N-abundance can be reproduced by {\sc Bonnsai} assuming moderate rotation ($\vrot\sim170\,\kms$) and relatively high inclination ($\sin i \sim 0.8$). In the case of VFTS~226 we were not able to extract measurements from the \ion{N}{iii}\,$\lambda$4511-4515 lines, although the line profiles resulting from measurements to the (weak) \ion{N}{iii}\,$\lambda$4379, 4097 lines are consistent with the data. The \ion{N}{ii}\,$\lambda3995$ line is (weakly) predicted by the models, but not present in the data. To reproduce its N-abundance, {\sc Bonnsai} predicts VFTS~226 is a rapidly rotating star seen almost pole on. VFTS~306 is part of the low gravity sample. The line profiles of \ion{N}{iii}\,$\lambda$4195, 4379, 4515, 4535 resulting from our measurement slightly underpredict the observed ones (by $\lesssim1\%$ of the continuum level), however the models corresponding to the 1$\sigma$ error on \nabun\ typically span the observed data. The \ion{N}{iii}\,$\lambda4097$ line is underpredicted by 10\% of the continuum and is not used for the abundance measurement.

{\it Confirmed binaries:} VFTS~064, 171, 332, 333 and 440 are confirmed by Almeida et al. (in prep.) as long-period ($\sim$\,$10^{3}\,\mathrm{d}$) SB1 systems. VFTS~093 was found to be a $250\,\mathrm{d}$ binary.  VFTS~399 has been extensively discussed by \citet{clark2015} and is potentially an X-ray binary. Interestingly, for VFTS~333, 399 and 440, our \nabun\ measurement agrees within 1$\sigma$ with the abundance predicted by {\sc Bonnsai}, based on their evolutionary state. VFTS~064 is N-enriched and shows \ion{N}{iii} emission at $4640\,\AA$, which is underpredicted in our best-fit model by $\sim$3\% of the continuum. VFTS~093, 171, 332 and 333 are late LC III-II stars with mild N-enrichment ($\nabun$ in the range $7.3$-$7.6$). In all four cases the profile of \ion{N}{iii}\,$\lambda4097$ is underpredicted. As these are long-period systems, their evolution is possibly unaffected by binary interaction (unless they are, e.g., former triples). In the case of VFTS~399, due to blending of the \ion{N}{iii}\,$\lambda$4511-4515-4518 complex, we extract an \Weq\ measurement by direct integration. For VFTS~440, Ram\'{i}rez-Agudelo et al. (subm.) report an unsatisfactory fit and parameters must be taken with caution. These 7 sources are kept in the paper as the VFTS sample of presumed single stars is constructed by well-defined criteria \citep[see][for a discussion]{sana2013}, benefiting binary population synthesis calculations. However, they are excluded from the comparison in Sect.~\ref{sec:popsyn_lowg}.

{\it Variable stars:} Almeida et al. (in prep.) find evidence for periodicity in the case of VFTS~259 and 764, however no radial velocity curve could be constructed that would indicate a binary nature. Quasi-periodic oscillations (e.g., pulsations) are possibly the cause of the (small) radial velocity variations. Both sources are classified as LC I, VFTS~259 is the only O-supergiant of intermediate spectral subtype in the VFTS. It has a complicated nitrogen spectrum, as the \ion{N}{iii} complex at $4515\,\AA$ transitions from absorption into emission at these temperatures (see Fig.~\ref{fig:TMNgrid}). In addition, the \ion{N}{iv}\,$\lambda4058$ line is predicted in the models but not present in the data. VFTS~764 is tagged as N-strong by \citet{walborn2014}, in agreement with our quantitative results. It is a relatively evolved star ($\logg_c = 2.90$). The \ion{N}{ii}\,$\lambda3995$ line is not reproduced in parallel to the \ion{N}{iii} lines (compared to the data, it is only weakly present in the models). We exclude it from the abundance measurement, as the \ion{N}{iii} lines yield consistent results. It remains to be investigated whether the irregular radial velocity variability of VFTS~259 and 764 offers an indication to the nature of their nitrogen enrichment.

{\it Confirmed single stars:} In the case of VFTS~087, 267, 178 and 819, Almeida et al. (in prep.) find no evidence for binarity. VFTS~087 and 178 are evolved supergiants of intermediate mass ($\sim30$-$35\,\msun$). In current evolutionary tracks, such stars are not expected to have lost a significant part of their envelope in prior phases. VFTS~267 is an early-type star with moderate enrichment, that is compatible with the nitrogen content predicted by {\sc Bonnsai}. For this star we were not able to extract a reliable \Weq\ measurement from its (noisy) \ion{N}{v}\,$\lambda4619$ line. The other two lines (\ion{N}{iv}\,$\lambda4058$ and \ion{N}{v}\,$\lambda4603$) are in excellent agreement, resulting in a very small error bar ($\sigma_\mathrm{N}=0.03$). Taking into account the error on the atmospheric parameters, Ram\'{i}rez-Agudelo et al. (subm.) find a more realistic 1$\sigma$ uncertainty of $\sim0.2\,\mathrm{dex}$ (see also Sect.~\ref{sec:limitations}). However, the line profiles of our best-fit model are shifted by $\sim1\,\AA$ and $~\sim2\,\AA$ for the \ion{N}{iv} and \ion{N}{v} profile respectively.  VFTS~819 is an ON star of moderate evolutionary mass ($22\,\msun$) and age ($\sim 2\,\mathrm{Myr}$). This source is very enriched in He as well as in N, and, given the high degree of abundance anomaly, is possibly a post-interaction binary. Compared to the tracks of \citet{brott2011a}, it could also be a quasi chemically-homogeneous star rotating initially at $\sim 500\,\kms$, seen almost pole-on.

{\it Other enriched sources:} The remainder of the stars in our sample was not included in the radial velocity follow-up survey. We comment here on the particularly enriched sources. VFTS~151, 180, 518, 599 and 669 show small but significant radial velocity variations, but were not observed by Almeida et al. (in prep.) for reasons related to fibre placement constraints. For VFTS~253, 466, 513, 566, 753 and 807, \citet{sana2013} did not find significant radial velocity variations. We first discuss the early-type stars.

VFTS~180, 518, 566 and 599 are O2-3 stars with high mass-loss rates and high-degrees of N-enrichment.  This could, in principle, point to their envelopes being stripped by their strong outflows, revealing chemically enriched layers.  This process might be ongoing in VFTS 180 that follows the empirical correlation between surface helium abundance and  $\log(\dot{M} / M)$ reported by \citet[][see also Sect.~\ref{sec:windstripping}]{bestenlehner}, but is not expected for the other sources, since these appear to be rather young objects.

Regarding their individual analyses, the early supergiant VFTS~180 shows very strong He- and N-emission. 
For this star, we find it challenging to reproduce the \ion{N}{v} lines in parallel to lines of other ionisation stages.  \ion{N}{v}\,$\lambda4603$ presents a P Cygni profile, and a \Weq\ measurement is not appropriate. We excluded \ion{N}{v}\,$\lambda4619$ as well because it is not well reproduced using the parameters provided by Ram\'{i}rez-Agudelo et al. (subm.). If we include this line the resulting abundance measurement does not change significantly (yielding $\nabun=8.70\pm0.40$), due to the relatively large error on its individual abundance measurement ($\sigma_i=0.10$). The continuum normalisation to the right of the \ion{N}{iii}\,$\lambda4630-4634-4640$ complex is shifted by $\sim 3\%$ upwards, to the left it is shifted by $\sim 1.5\%$. Although the exact degree of N-enrichment is difficult to constrain, it is certainly enriched. 
For VFTS~518 the best fit model underpredicts the \ion{N}{v} lines (at a level of $\sim$2\% of the continuum), but fits well the \ion{N}{iii}\,$\lambda$4634-4640-4641 and \ion{N}{iv}\,$\lambda$4058 lines. This source is also enriched in He. In the cases of VFTS~566 and 599 the best-fit model of the \ion{N}{v} lines is slightly shifted (by $\sim0.5\,\AA$) and underpredicted compared to the data (by $\sim2\%$ of the continuum).

VFTS 151 is a spatially resolved triple within the VFTS fibre producing a composite spectrum with a large range of types (see Table A.1 of Walborn et al. 2014). 
We are unable to reproduce the \ion{N}{iv}\,$\lambda$4058 line and underpredict the \ion{N}{iii}\,$\lambda$4634-4640-4641 complex by 2\% of the continuum, while the \ion{N}{iii}\,$\lambda$4511-4515-4518 complex, predicted by the best model, appears absent. 
This star does not make part of the low gravity sample. For VFTS~253 and 513, N-lines are present that have widths consistent with the resolution limit of the survey, even though Ram\'{i}rez-Agudelo et al. (subm.) estimate $\vsini\sim100\,\kms$ for both sources. Test models with \vsini\ values below the resolution limit provide a satisfactory fit. However, it cannot be excluded that these lines originate from an undetected companion. The fibre of VFTS~513 could be contaminated, but for VFTS~253 this seems not the case. Both stars have rotation-corrected gravities in excess of 4, and VFTS~253 is one of the stars with Si-weak lines \citep[see][]{walborn2014}. VFTS~466 and 753 are otherwise unremarkable late giants, with N-abundances of $\sim7.6$. VFTS~669 is a late supergiant. It shows emission at \ion{N}{iii}\,$\lambda$4640 which is not reproduced by our best fit model. Nevertheless, good agreement is found with the N-abundance determined by Ram\'{i}rez-Agudelo et al. (subm.). VFTS~807 is tagged with N-strong by \citet{walborn2014}, in conformity with our N-measurement. It is a moderately evolved, but not so massive star $M_\mathrm{evol}=18.8\,\msun$). The origin of the observed N-enrichment in aforementioned stars remains to be investigated.

Finally we note that VFTS~016, 064, 171, 180, 259, 267, 333, 518, 566, 599, 664 and 669 were also studied by \citet{bestenlehner}. They provide no nitrogen abundance measurements, but rate stars as nitrogen normal, enhanced, or in-between. These qualitative measures are in general agreement with our quantitative results.

\section{Additional examples of line fits}
\label{app:examplesenriched}

\begin{figure*}[t!]
\centering
\includegraphics[width=\textwidth]{\figpath 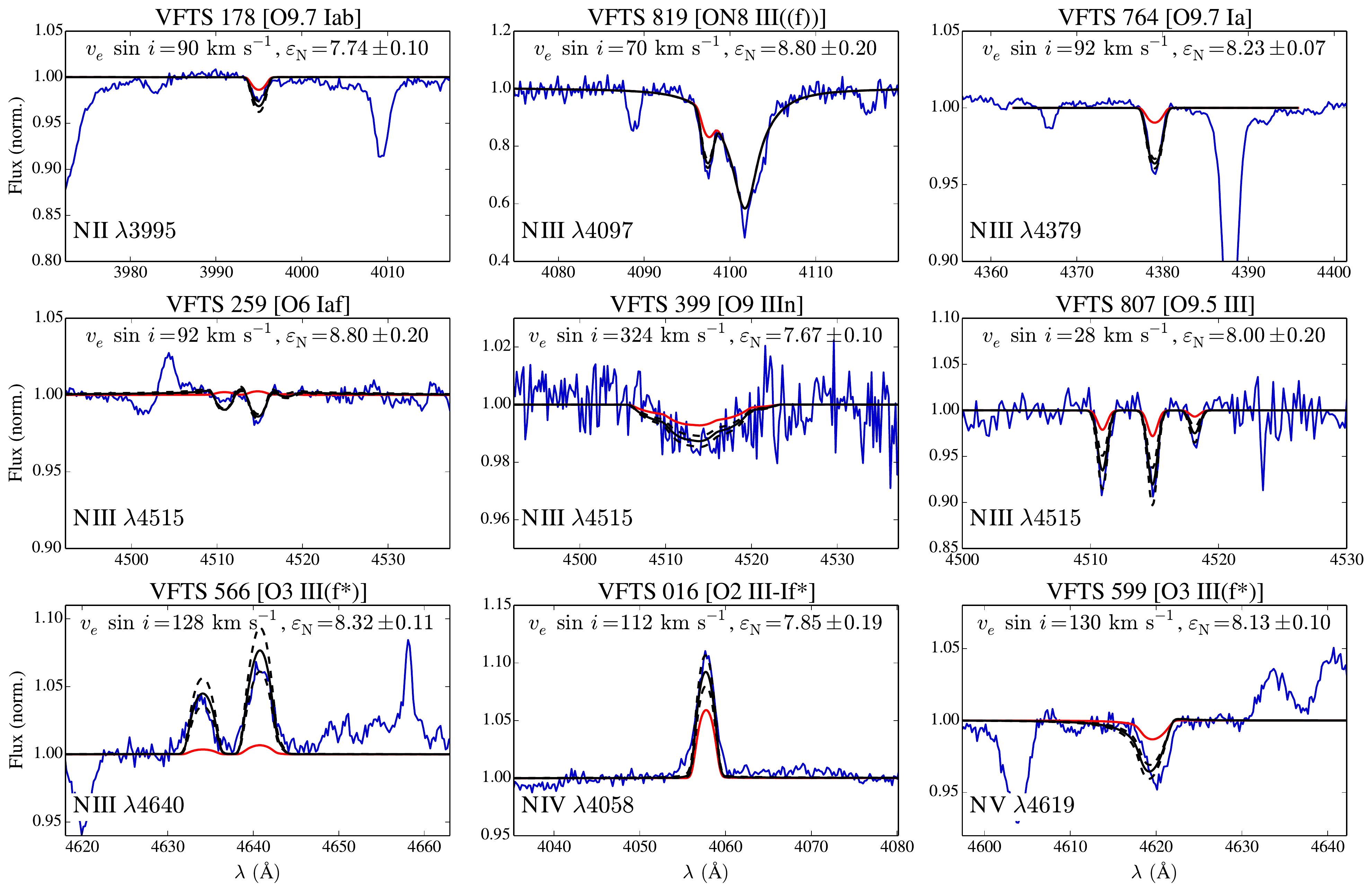}
\caption{Examples of line fits for enriched stars, ranging over spectral types and nitrogen ionisation stage used. The observed spectrum is shown in blue, the best fit model in black (the measured $\nabun$ is annotated in the panel), models representing the upper and lower 1$\sigma$ error in dashed black, and a {\sc fastwind} model with $\nabun=7.33$ in red. Only one of the lines used is shown for each source. All sources need a significant nitrogen enrichment in order to reproduce the observed profiles.}
\label{fig:examplesenriched}
\end{figure*}

\end{appendix}

\end{document}